# Error Exponents for the Gaussian Channel with Active Noisy Feedback

Young-Han Kim     Amos Lapidoth     Tsachy Weissman

November 17, 2018

**Abstract**

We study the best exponential decay in the blocklength of the probability of error that can be achieved in the transmission of a single bit over the Gaussian channel with an active noisy Gaussian feedback link. We impose an *expected* block power constraint on the forward link and study both *almost-sure* and *expected* block power constraints on the feedback link. In both cases the best achievable error exponents are finite and grow approximately proportionally to the larger between the signal-to-noise ratios on the forward and feedback links. The error exponents under almost-sure block power constraints are typically strictly smaller than under expected constraints. Some of the results extend to communication at arbitrary rates below capacity and to general discrete memoryless channels.

## 1 Introduction

This paper studies error exponents for the Gaussian channel with noisy feedback. Unlike our previous work, which focused on *passive* feedback [1], [2], [3], here we focus on *active* feedback. Thus, the time-$k$ symbol $U_k$ fed to the feedback channel need not be the time-$k$ received symbol $Y_k$: it can be a function of $Y_k$ and of the previous received symbols $Y_1, \ldots, Y_{k-1}$. As in our previous work, we consider only transmission schemes of a deterministic blocklength $n$. (Random transmission times for discrete memoryless channels with active feedback are discussed in [4].) And, although some of our results extend to more general models, we focus on the Gaussian model where both the forward channel and the feedback channel are additive white Gaussian noise channels. To simplify the analysis we focus on the case where the message to be transmitted is binary, i.e., takes on the values 0 and 1 equiprobably (but see (13) which is applicable to all rates of communication between zero and capacity). Our communication scheme is depicted in Figure 1.

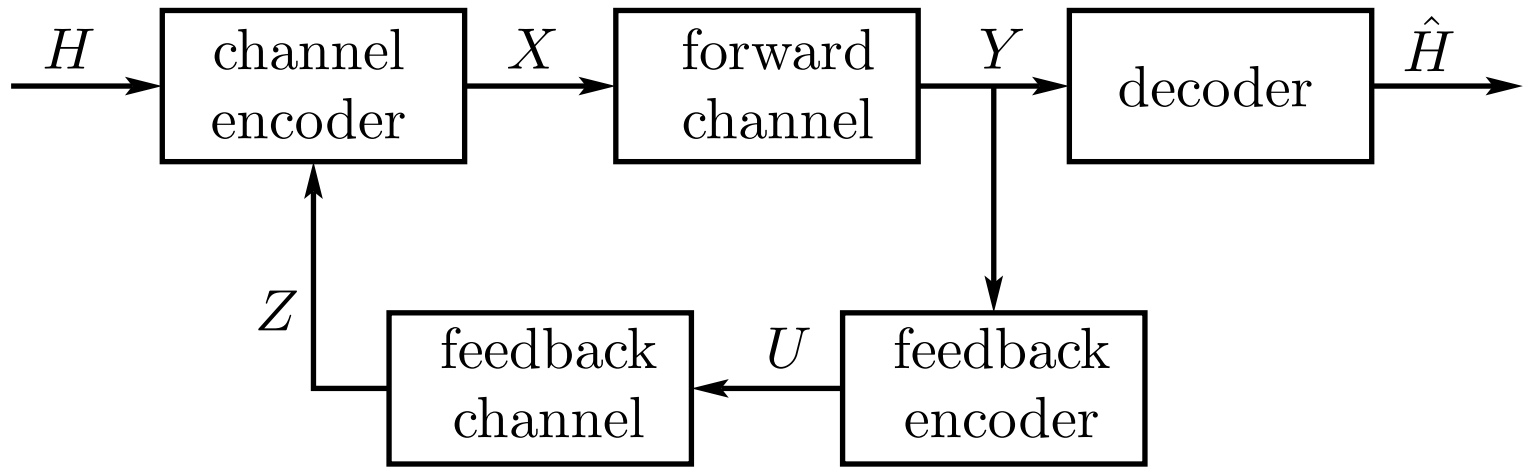

Figure 1: The Gaussian channel with a coded noisy feedback link.

Critical to our analysis is the precise nature of the power constraints that are imposed on the forward and feedback channels. On the forward channel we impose an *expected block power constraint*, where the time-average of the squared channel inputs is a random variable (whose realization may depend on the message and on the realization of the forward and feedback channels) whose expectation (over messages and the noise sequences on the forward and feedback channels) is upper-bounded by some fixed (deterministic) positive constant $\mathsf{P}$; see (8) ahead. For the feedback link we consider two types of power constraints: an expected block power constraint ((9) ahead) and an *almost-sure block power constraint* ((10)). In the latter, the time-average of the squared inputs to the feedback channel must not exceed $\mathsf{P}_{\mathrm{FB}}$ *irrespective of the message and of the channel realizations.* Clearly, an almost-sure power constraint is more restrictive than an expected power constraint.

We do not consider an almost-sure block power constraint on the forward channel because under this constraint even a noise-free feedback link does not improve the two-codewords error exponent [5], [6].

Our main result is that—although a noise-free feedback link allows the probability of error to decay faster than exponentially in $n$ [7] [8] [9]—if the feedback link is noisy the probability of error cannot decay faster than exponentially. This is true even if we only impose an expected block power constraint on the feedback link. Moreover, we provide upper and lower bounds on the best achievable exponent both for expected and almost-sure block power constraints. At high signal-to-noise ratios (SNRs) on the feedback link, the error exponents in both cases grow as an affine function of the SNR. A more formal statement of the results will be give in Section 2 once we have formalized the problem's statement.

# 2 The Problem Statement and Main Results

We consider the transmission of a single bit $H$, where $H$ takes on the values 0 and 1 equiprobably. Let the sets $\mathcal{X}$, $\mathcal{Y}$, $\mathcal{U}$, and $\mathcal{Z}$ all be the reals. A blocklength-$n$ code for transmitting $H$ over our channel consists of a forward-channel encoding rule, a feedback-channel encoding rule, and a decoder as described next. A forward-channel encoding rule is specified by $n$ functions[1] $f_1, \ldots, f_n$, where

$$f_k \colon \{0,1\} \times \mathcal{Z}^{k-1} \to \mathcal{X}, \quad k = 1, \ldots n. \tag{1}$$

It is understood that the time-$k$ channel input $X_k$ is computed according to the rule

$$X_k = f_k\big(H, Z^{k-1}\big), \quad k = 1, \ldots, n, \tag{2}$$

where we use $A^\ell$ to denote $A_1, \ldots, A_\ell$ and where, for convenience, we set

$$Z_0 = 0. \tag{3}$$

The feedback-channel encoding rule is a collection of $n$ functions $g_1, \ldots, g_n$, where

$$g_k \colon \mathcal{Y}^k \to \mathcal{U}, \quad k = 1, \ldots, n. \tag{4}$$

It is understood that the symbol $U_k$ that is fed to the feedback channel at time $k$ is given by

$$U_k = g_k\big(Y^k\big), \quad k = 1, \ldots, n. \tag{5}$$

(The special case where $g_k(Y^k)$ is $Y_k$ corresponds to passive—also knows as "uncoded" or "symbol-by-symbol"–feedback.)

A decoder $\boldsymbol{\phi}$ is a decision rule for guessing $H$ based on $Y^n$. Thus,

$$\boldsymbol{\phi} \colon \mathcal{Y}^n \to \{0,1\}. \tag{6}$$

[1]All functions from $\mathbb{R}$ to $\mathbb{R}$ in this paper are assumed to be Borel Measurable.

We denote the decision regions by $\mathcal{D}_0$ and $\mathcal{D}_1$ so

$$\mathcal{D}_\nu = \bigl\{\mathbf{y} \in \mathcal{Y}^n : \phi(\mathbf{y}) = \nu\bigr\}, \quad \nu = 0, 1, \tag{7}$$

where we use $\mathbf{a}$ to denote the $n$-tuple $(a_1, \dots, a_n)$.

The communication system that we consider operates as follows. The message, along with the forward and backward channel noise components, $H, N_1, \dots, N_n, V_1, \dots, V_n$, are independent random variables, where $N_k \sim \mathcal{N}(0, \sigma^2)$ and $V_k \sim \mathcal{N}(0, \sigma^2_{\mathrm{FB}})$ for every $k \in \{1, \dots, n\}$. We assume throughout that $\sigma$ and $\sigma_{\mathrm{FB}}$ are strictly positive. At time $k$, the input $X_k$ to the forward channel is generated according to (2). This input is corrupted by the forward channel noise, yielding the forward-channel output $Y_k = X_k + N_k$. The feedback-channel encoder now computes the symbol $U_k$ from $Y^k$ according to (5). The symbol $U_k$, which forms the time-$k$ input to the feedback channel is corrupted by the feedback-channel noise to yield $Z_k = U_k + V_k$ at the output of that channel. The conditional density $w(y_k|x_k)$ of $Y_k$ given $X_k$ is thus

$$w(y_k|x_k) = \frac{1}{\sqrt{2\pi\sigma^2}}\, e^{-\frac{(y_k - x_k)^2}{2\sigma^2}}, \quad x_k, y_k \in \mathbb{R},$$

and the conditional density $w_{\mathrm{FB}}(z_k|u_k)$ of $Z_k$ given $U_k$ is

$$w_{\mathrm{FB}}(z_k|u_k) = \frac{1}{\sqrt{2\pi\sigma^2_{\mathrm{FB}}}}\, e^{-\frac{(z_k - u_k)^2}{2\sigma^2_{\mathrm{FB}}}}, \quad z_k, u_k \in \mathbb{R}.$$

We only consider forward-channel encoding rules that satisfy the expected block power constraint

$$\mathsf{E}\Biggl[\sum_{k=1}^{n} f_k^2\bigl(H, Z^{k-1}\bigr)\Biggr] \le n\mathsf{P}, \tag{8}$$

where $\mathsf{P} > 0$ is some given constant designating the allowed average power (per transmission) on the forward-channel.

For the feedback-channel encoding rules we consider two types of power constraints. An almost-sure block power constraint

$$\sum_{k=1}^{n} g_k^2\bigl(Y^k\bigr) \le n\mathsf{P}_{\mathrm{FB}}, \quad \mathbf{Y} \in \mathcal{Y}^n, \tag{9}$$

and an expected block power constraint

$$\mathsf{E}\left[\sum_{k=1}^{n} g_k^2\big(Y^k\big)\right] \le n\mathsf{P}_{\mathrm{FB}}. \tag{10}$$

In both cases we assume that $\mathsf{P}_{\mathrm{FB}}$ is strictly positive. (The case where $\mathsf{P}_{\mathrm{FB}} = 0$ corresponds to the no-feedback case.) We can now present our main results.
**Almost-Sure Block Power Constraints:** Let $p_{\mathrm{e}}^{\mathrm{a.s.}}\big(\mathsf{P}/\sigma^2, \mathsf{P}_{\mathrm{FB}}/\sigma_{\mathrm{FB}}^2, n\big)$ denote the least probability of error that can be achieved by a blocklength-$n$ coding scheme subject to the almost-sure constraint (9). We show in Section 4 that

$$\varlimsup_{n\to\infty} -\frac{1}{n}\log p_{\mathrm{e}}^{\mathrm{a.s.}}\big(\mathsf{P}/\sigma^2, \mathsf{P}_{\mathrm{FB}}/\sigma_{\mathrm{FB}}^2, n\big) \le \frac{\mathsf{P}}{2\sigma^2} + \frac{2\sqrt{(\mathsf{P}_{\mathrm{FB}}+\sigma_{\mathrm{FB}}^2)\mathsf{P}_{\mathrm{FB}}}}{\sigma_{\mathrm{FB}}^2}, \tag{11}$$

and we present in Section 5 a sequence of codes that proves that

$$\varliminf_{n\to\infty} -\frac{1}{n}\log p_{\mathrm{e}}^{\mathrm{a.s.}}\big(\mathsf{P}/\sigma^2, \mathsf{P}_{\mathrm{FB}}/\sigma_{\mathrm{FB}}^2, n\big) \ge \frac{\mathsf{P}}{2\sigma^2} + \frac{2\mathsf{P}_{\mathrm{FB}}}{\sigma_{\mathrm{FB}}^2}. \tag{12}$$

It is shown in Section 4 that (11) generalizes to the case where there are more than two codewords. If we denote by $R$ the rate of communication, i.e., the ratio of the logarithm of the number of messages to the block length, and if we denote by $E_{\mathrm{FB}}^{\mathrm{a.s.}}(R)$ the best achievable error exponent then

$$E_{\mathrm{FB}}^{\mathrm{a.s.}}(R) \le E_{\mathrm{NoFB}}(R) + \frac{2\sqrt{(\mathsf{P}_{\mathrm{FB}}+\sigma_{\mathrm{FB}}^2)\mathsf{P}_{\mathrm{FB}}}}{\sigma_{\mathrm{FB}}^2}, \tag{13}$$

where $E_{\mathrm{NoFB}}(R)$ is the reliability function of the forward channel in the absence of feedback.
**Expected Block Power Constraints:** Let $p_{\mathrm{e}}^{\exp}\big(\mathsf{P}/\sigma^2, \mathsf{P}_{\mathrm{FB}}/\sigma_{\mathrm{FB}}^2, n\big)$ denote the least probability of error that can be achieved by a blocklength-$n$ coding scheme subject to the expected block power constraint (10). We show in Section 6 that[2]

$$\varlimsup_{n\to\infty} -\frac{1}{n}\log p_{\mathrm{e}}^{\exp}\big(\mathsf{P}/\sigma^2, \mathsf{P}_{\mathrm{FB}}/\sigma_{\mathrm{FB}}^2, n\big) \le \frac{\left(\sqrt{\mathsf{P}+\sigma^2}+\sqrt{\mathsf{P}}\right)^2}{\sigma^2} + \frac{\left(\sqrt{\mathsf{P}_{\mathrm{FB}}+\sigma_{\mathrm{FB}}^2}+\sqrt{\mathsf{P}_{\mathrm{FB}}}\right)^2}{\sigma_{\mathrm{FB}}^2}, \tag{14}$$

[2]Section 6 also presents a tighter bound than (14).

and we present in Section 7 a sequence of codes that achieves

$$\varliminf_{n\to\infty} -\frac{1}{n}\log p_{\mathrm{e}}^{\exp}\bigl(\mathsf{P}/\sigma^2, \mathsf{P}_{\mathrm{FB}}/\sigma_{\mathrm{FB}}^2, n\bigr) \geq \frac{2\mathsf{P}}{\sigma^2} + \frac{2\mathsf{P}_{\mathrm{FB}}}{\sigma_{\mathrm{FB}}^2}. \tag{15}$$

# 3 Some Preliminaries

Fix some code, and consider its use for transmitting the bit $H$ over our channel. Let $\mathbf{X}$ denote the forward-channel inputs that result from such use through (2). Similarly define the forward-channel output sequence $\mathbf{Y}$, the feedback-channel inputs by $\mathbf{U}$, and the feedback-channel outputs by $\mathbf{Z}$. Let $\mathsf{P}$ denote the joint law of $\mathbf{X}, \mathbf{Y}, \mathbf{U}, \mathbf{Z}$ induced by the coding scheme, and let $p(\mathbf{y}, \mathbf{z})$ denote the joint density of $\mathbf{Y}, \mathbf{Z}$. Let the conditional versions of the joint law and density given $H = 0$ be denoted by $\mathsf{P}_0$ and $p_0(\mathbf{y}, \mathbf{z})$. Similarly we use $\mathsf{P}_1$ and $p_1(\mathbf{y}, \mathbf{z})$ when the conditioning is on $H = 1$. Thus

$$\begin{aligned} p_\nu(\mathbf{y}, \mathbf{z}) &= \prod_{k=1}^{n} p_\nu\bigl(y_k, z_k | y^{k-1}, z^{k-1}\bigr) \\ &= \prod_{k=1}^{n} \Bigl(p_\nu\bigl(y_k | y^{k-1}, z^{k-1}\bigr) p_\nu\bigl(z_k | y^k, z^{k-1}\bigr)\Bigr) \\ &= \prod_{k=1}^{n} w\Bigl(y_k \big| f_k\bigl(\nu, z^{k-1}\bigr)\Bigr) \prod_{k=1}^{n} w_{\mathrm{FB}}\Bigl(z_k \big| g_k\bigl(y^k\bigr)\Bigr), \quad \nu = 0, 1, \end{aligned} \tag{16}$$

where the last equality follows from our assumption that both the forward and feedback channels are memoryless and that the noise components of these channels are independent. Since $H$ takes on the values 0 and 1 equiprobably we have

$$p(\mathbf{y}, \mathbf{z}) = \frac{1}{2}\, p_0(\mathbf{y}, \mathbf{z}) + \frac{1}{2}\, p_1(\mathbf{y}, \mathbf{z}), \tag{17}$$

and the probability of error $\mathsf{P}(\text{error})$ can be expressed as

$$\begin{aligned} \mathsf{P}(\text{error}) &= \frac{1}{2}\, \mathsf{P}_0(\text{error}) + \frac{1}{2}\, \mathsf{P}_1(\text{error}) \qquad (18) \\ &= \frac{1}{2}\, \mathsf{P}_0\bigl(\mathbf{Y} \in \mathcal{D}_1\bigr) + \frac{1}{2}\, \mathsf{P}_1\bigl(\mathbf{Y} \in \mathcal{D}_0\bigr). \qquad (19) \end{aligned}$$

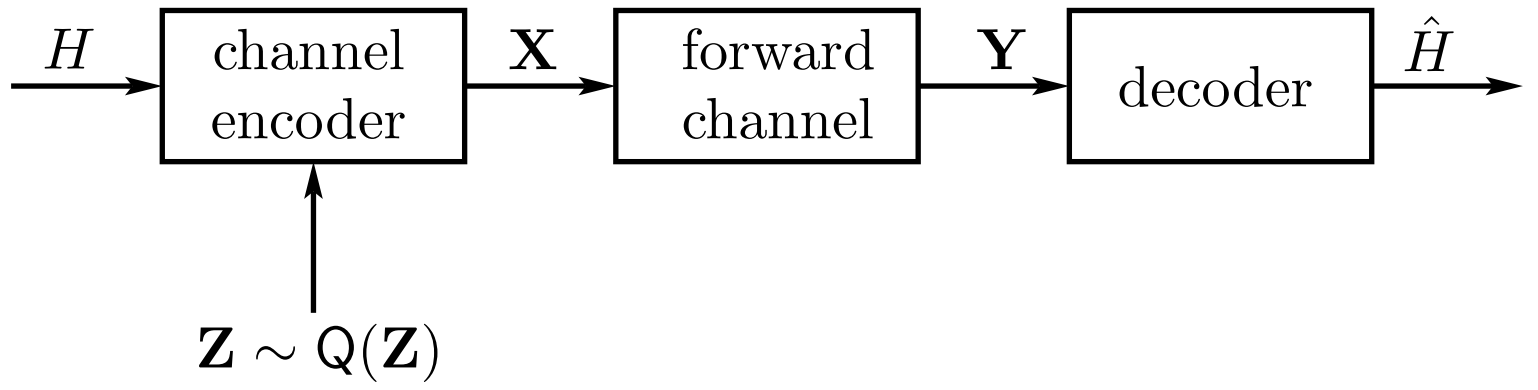

Figure 2: The new channel in which $\mathbf{Z}$ is generated at random.

# 4 Inachievability under A.S. Power Constraints

The derivation of (11) is based on a new reference channel that we define next.

## 4.1 A Reference Channel

We consider a new channel in which the forward-channel encoder, feedback-channel encoder, and channel decoder are as above, but in which the sequence $\mathbf{z}$ that is fed to the forward-channel encoder is generated independently of $H$ according to some law $\mathsf{Q}$; see Figure 2. Later we shall assume that this law is the marginal law $\mathsf{P}(\mathbf{Z})$ that is induced by the original channel and code. We denote the density of this law by $q(\mathbf{z})$. Thus, the new channel operates like the original channel except that the sequence $\mathbf{z}$ that is fed to the channel encoder is not the output of the feedback channel but rather a randomly generated sequence drawn independently of $H$ according to $q(\mathbf{z})$. We denote by $\mathsf{Q}$ the joint distribution of $H$, $\mathbf{X}$, $\mathbf{Y}$, and $\mathbf{Z}$ that results when the original forward-channel encoder (2) is applied as in Figure 2. Thus, if we factorize the marginal law of $\mathsf{Q}$ on $\mathcal{Z}^n$ as

$$\mathsf{Q}(\mathbf{z}) = \prod_{k=1}^{n} \mathsf{Q}_{Z_k|Z^{k-1}}\big(z_k\big|z^{k-1}\big) \tag{20}$$

then

$$\mathsf{Q}\big(Z_k = z_k\big|Z^{k-1} = z^{k-1}, Y^k = y^k, H = \nu\big) = \mathsf{Q}_{Z_k|Z^{k-1}}\big(z_k\big|z^{k-1}\big). \tag{21}$$

(Under $\mathsf{Q}$ the sequence $\mathbf{Z}$ is generated "without looking at $\mathbf{Y}$." Nevertheless, $\mathbf{Z}$ and $\mathbf{Y}$ are not necessarily independent under $\mathsf{Q}$ because $\mathbf{Y}$ is influenced via the forward-channel by $\mathbf{X}$, which in turn is influenced by $\mathbf{Z}$ via the forward-channel encoder.)

The conditional laws of $\mathsf{Q}$ conditional on $H = 0$ and $H = 1$ are denoted by $\mathsf{Q}_0$, $\mathsf{Q}_1$. Similarly, the unconditional and conditional joint densities of $\mathbf{Y}, \mathbf{Z}$ are denoted by $q(\mathbf{y}, \mathbf{z})$, $q_0(\mathbf{y}, \mathbf{z})$, and $q_1(\mathbf{y}, \mathbf{z})$.

Note that according to $\mathsf{Q}$ the random sequence $\mathbf{Z}$ is generated independently of the hypothesis $H$, so

$$q_0(\mathbf{z}) = q_1(\mathbf{z}) = q(\mathbf{z}). \tag{22}$$

Also,

$$\begin{aligned} q_\nu(\mathbf{y}, \mathbf{z}) &= \prod_{k=1}^{n} q_\nu\big((y_k, z_k)\big|y^{k-1}, z^{k-1}\big) \\ &= \prod_{k=1}^{n} q_\nu\big(z_k\big|y^k, z^{k-1}\big) \prod_{k=1}^{n} q_\nu\big(y_k\big|y^{k-1}, z^{k-1}\big) \\ &= \prod_{k=1}^{n} q\big(z_k\big|z^{k-1}\big) \prod_{k=1}^{n} w\big(y_k\big|f_k(\nu, z^{k-1})\big) \\ &= q(\mathbf{z}) \prod_{k=1}^{n} w\big(y_k\big|f_k(\nu, z^{k-1})\big), \quad \nu = 0, 1, \end{aligned} \tag{23}$$

where the first equality follows from the chain rule; the second from another application of the chain rule; the third from (21); and the fourth from the chain rule.

The importance of the reference law is that the set-up of Figure 2 is no better than the no-feedback set-up. And since antipodal signaling is optimal in the absence of feedback,

$$\mathsf{Q}(\text{error}|\mathbf{z}) \geq \mathcal{Q}\left(\sqrt{\frac{\sum_{k=1}^{n} f_k^2(0, z^{k-1}) + \sum_{k=1}^{n} f_k^2(1, z^{k-1})}{2\sigma^2}}\right), \quad \mathbf{z} \in \mathcal{Z}^n. \tag{24}$$

## 4.2 Relating $\mathsf{Q}$ and $\mathsf{P}$

We next try to relate the original law $\mathsf{P}$ and the reference law $\mathsf{Q}$. By (16) and (23) it follows that

$$p_\nu(\mathbf{y}, \mathbf{z}) = r(\mathbf{y}, \mathbf{z})\, q_\nu(\mathbf{y}, \mathbf{z}), \quad \nu = 0, 1, \tag{25}$$

where

$$r(\mathbf{y},\mathbf{z}) \triangleq \frac{\prod_{k=1}^{n} w_{\mathrm{FB}}\big(z_k \big| g_k(y^k)\big)}{q(\mathbf{z})}. \tag{26}$$

In order to lower-bound $\mathsf{P}(\text{error})$ in terms of $\mathsf{Q}(\text{error})$, we need a lower bound on $r(\mathbf{y},\mathbf{z})$. We derive one for the choice of $\mathsf{Q}(\mathbf{z})$ as $\mathsf{P}(\mathbf{z})$. Henceforth we thus assume

$$q(\mathbf{z}) = p(\mathbf{z}), \quad \mathbf{z} \in \mathcal{Z}^n. \tag{27}$$

It follows from (26) and (27) that in order to lower-bound $r(\mathbf{y},\mathbf{z})$ we need to upper-bound $p(\mathbf{z})$. To that end, we first use (16) to express $p_\nu(\mathbf{y},\mathbf{z})$ as

$$p_\nu(\mathbf{y},\mathbf{z}) = \prod_{k=1}^{n} w_{\mathrm{FB}}(z_k|u_k) \prod_{k=1}^{n} p_\nu\big(y_k \big| z^{k-1}, y^{k-1}\big), \tag{28a}$$

where

$$u_k = g_k(y^{k-1}), \quad k = 1, \ldots, n. \tag{28b}$$

For a fixed $\mathbf{z} \in \mathcal{Z}^n$, the second product on the RHS of (28a) is a probability density function that integrates to one over the set of all sequences $\mathbf{y} \in \mathcal{Y}^n$ that, via (28b), induce $\mathbf{u}$ sequences satisfying

$$\frac{1}{n}\sum_{k=1}^{n} u_k^2 \le \mathsf{P}_{\mathrm{FB}}. \tag{29}$$

Since the average is upper-bounded by the maximum, we obtain

$$p_\nu(\mathbf{z}) = \int p_\nu(\mathbf{y},\mathbf{z})\,\mathrm{d}\mathbf{y} \tag{30}$$

$$\le \max_{\mathbf{u}} \prod_{k=1}^{n} w_{\mathrm{FB}}(z_k|u_k), \tag{31}$$

where the maximum is over all the sequences $\mathbf{u} \in \mathcal{U}^n$ that satisfy (29).

Using the explicit form of the Gaussian distribution and the Cauchy-Schwarz Inequality we obtain

$$\begin{aligned} p_\nu(\mathbf{z}) &\le \max_{\|\mathbf{u}\|^2 \le n\mathsf{P}_{\mathrm{FB}}} \big(2\pi\sigma_{\mathrm{FB}}^2\big)^{-n/2} \exp\left(-\frac{1}{2\sigma_{\mathrm{FB}}^2}\big(\|\mathbf{z}\|^2 - 2\,\langle \mathbf{z},\mathbf{u}\rangle_{\mathrm{E}} + \|\mathbf{u}\|^2\big)\right) \\ &\le \max_{\|\mathbf{u}\|^2 \le n\mathsf{P}_{\mathrm{FB}}} \big(2\pi\sigma_{\mathrm{FB}}^2\big)^{-n/2} \exp\left(-\frac{1}{2\sigma_{\mathrm{FB}}^2}\big(\|\mathbf{z}\|^2 - 2\,\|\mathbf{z}\|\,\|\mathbf{u}\| + \|\mathbf{u}\|^2\big)\right) \end{aligned}$$

$$= \max_{\|\mathbf{u}\|^2 \le n\mathsf{P}_{\mathrm{FB}}} \left(2\pi\sigma_{\mathrm{FB}}^2\right)^{-n/2} \exp\left(-\frac{1}{2\sigma_{\mathrm{FB}}^2}\left(\|\mathbf{z}\| - \|\mathbf{u}\|\right)^2\right)$$

$$= \left(2\pi\sigma_{\mathrm{FB}}^2\right)^{-n/2} \exp\left(-\frac{1}{2\sigma_{\mathrm{FB}}^2}\left(\left(\|\mathbf{z}\| - \sqrt{n\mathsf{P}_{\mathrm{FB}}}\right)^+\right)^2\right), \quad \nu = 0, 1, \quad (32)$$

where $\xi^+$ denotes the larger of zero and $\xi$, i.e., $\max\{0, \xi\}$. Since (32) holds for both $\nu = 0$ and $\nu = 1$, we can average over $\nu$ to obtain

$$p(\mathbf{z}) \le \left(2\pi\sigma_{\mathrm{FB}}^2\right)^{-n/2} \exp\left(-\frac{1}{2\sigma_{\mathrm{FB}}^2}\left(\left(\|\mathbf{z}\| - \sqrt{n\mathsf{P}_{\mathrm{FB}}}\right)^+\right)^2\right). \quad (33)$$

We continue to lower-bound $r(\mathbf{y}, \mathbf{z})$, which is defined in (26) by lower-bounding the numerator using (9). Again, using the explicit form of the Gaussian distribution and the Cauchy-Schwarz Inequality we obtain

$$\prod_{k=1}^{n} w_{\mathrm{FB}}\Big(z_k | g_k\big(y^k\big)\Big) \ge \left(2\pi\sigma_{\mathrm{FB}}^2\right)^{-n/2} \exp\left(-\frac{1}{2\sigma_{\mathrm{FB}}^2}\left(\|\mathbf{z}\| + \sqrt{n\mathsf{P}_{\mathrm{FB}}}\right)^2\right) \quad (34)$$

Combining (26), (27), (33), and (34) we obtain

$$r(\mathbf{y}, \mathbf{z}) \ge \exp\left(-\frac{2\,\|\mathbf{z}\|\,\sqrt{n\mathsf{P}_{\mathrm{FB}}}}{\sigma_{\mathrm{FB}}^2}\right), \quad \|\mathbf{z}\|^2 \ge n\mathsf{P}_{\mathrm{FB}}, \quad (35)$$

and

$$r(\mathbf{y}, \mathbf{z}) \ge \exp\left(-\frac{1}{2\sigma_{\mathrm{FB}}^2}\left(\|\mathbf{z}\| + \sqrt{n\mathsf{P}_{\mathrm{FB}}}\right)^2\right) \quad (36)$$

$$\ge \exp\left(-\frac{2\mathsf{P}_{\mathrm{FB}}}{\sigma_{\mathrm{FB}}^2} n\right), \quad \|\mathbf{z}\|^2 \le n\mathsf{P}_{\mathrm{FB}}. \quad (37)$$

These two inequalities can be combined to yield

$$r(\mathbf{y}, \mathbf{z}) \ge \exp\left(-\frac{2\max\{\|\mathbf{z}\|/\sqrt{n}, \sqrt{\mathsf{P}_{\mathrm{FB}}}\}\sqrt{\mathsf{P}_{\mathrm{FB}}}}{\sigma_{\mathrm{FB}}^2} n\right), \quad \mathbf{z} \in \mathcal{Z}^n. \quad (38)$$

Consequently, by (25)

$$p_\nu(\mathbf{y}, \mathbf{z}) \ge q_\nu(\mathbf{y}, \mathbf{z}) \exp\left(-\frac{2\max\{\|\mathbf{z}\|/\sqrt{n}, \sqrt{\mathsf{P}_{\mathrm{FB}}}\}\sqrt{\mathsf{P}_{\mathrm{FB}}}}{\sigma_{\mathrm{FB}}^2} n\right), \quad \mathbf{z} \in \mathcal{Z}^n. \quad (39)$$

Note that the RHS of (38) is monotonically decreasing in $\|\mathbf{z}\|$.

### 4.3 Lower-Bounding the Probability of Error

Fix some $\epsilon > 0$ and $\delta > 0$ smaller than 1/2. Define

$$\beta \triangleq \mathsf{P}_{\text{FB}} + \sigma_{\text{FB}}^2 + \delta. \tag{40}$$

Note that

$$\beta \geq \mathsf{P}_{\text{FB}}. \tag{41}$$

By the almost-sure block power constraint on the feedback channel (9), it follows that there exists some positive integer $n_0$ (which depends on the choice of $\epsilon$) such that for any code of blocklength $n$ exceeding $n_0$

$$\mathsf{P}(\mathcal{B}) \geq 1 - \epsilon, \tag{42}$$

where

$$\mathcal{B} \triangleq \{\mathbf{z} \in \mathcal{Z}^n : \|\mathbf{z}\|^2 \leq \beta n\}. \tag{43}$$

By Markov's inequality and by the expected block power constraint on the forward channel (8), it follows that for any code the subset $\mathcal{A}$ of $\mathcal{Z}^n$ consisting of those sequences that cause the transmitted symbols on the forward link to have an average power that does not exceed $\mathsf{P}/(1-2\epsilon)$, i.e., the set

$$\mathcal{A} \triangleq \left\{ \mathbf{z} \in \mathcal{Z}^n : \frac{1}{2n} \sum_{k=1}^{n} f_k^2(0, z^{k-1}) + \frac{1}{2n} \sum_{k=1}^{n} f_k^2(1, z^{k-1}) \leq \frac{\mathsf{P}}{1-2\epsilon} \right\} \tag{44}$$

satisfies

$$\mathsf{P}(\mathcal{A}) \geq 2\epsilon. \tag{45}$$

It follows from (45) and (42) that

$$\mathsf{P}(\mathcal{A} \cap \mathcal{B}) \geq \epsilon. \tag{46}$$

Recalling (41) and the monotonicity of the RHS of (38), we conclude that for $\mathbf{z} \in \mathcal{B}$ we can lower-bound $r(\mathbf{y}, \mathbf{z})$ by

$$r(\mathbf{y}, \mathbf{z}) \geq e^{-\alpha n}, \quad \mathbf{z} \in \mathcal{B}, \tag{47a}$$

where

$$\alpha \triangleq \frac{2\sqrt{\beta \mathsf{P}_{\text{FB}}}}{\sigma_{\text{FB}}^2}. \tag{47b}$$

We can now lower-bound the conditional probabilities of error as follows:

$$\begin{aligned}
\mathsf{P}_1(\mathbf{Y} \in \mathcal{D}_0) &= \int_{\mathcal{Z}^n} \mathsf{P}_1(\mathbf{Y} \in \mathcal{D}_0 | \mathbf{Z} = \mathbf{z})\, p_1(\mathbf{z})\, \mathrm{d}\mathbf{z} \\
&\geq \int_{\mathcal{A}\cap\mathcal{B}} \mathsf{P}_1(\mathbf{Y} \in \mathcal{D}_0 | \mathbf{Z} = \mathbf{z})\, p_1(\mathbf{z})\, \mathrm{d}\mathbf{z} \\
&= \int_{\mathcal{A}\cap\mathcal{B}} \int_{\mathcal{D}_0} p_1(\mathbf{y}, \mathbf{z})\, \mathrm{d}\mathbf{y}\, \mathrm{d}\mathbf{z} \\
&= \int_{\mathcal{A}\cap\mathcal{B}} \int_{\mathcal{Y}^n} \mathrm{I}\{\mathbf{y} \in \mathcal{D}_0\}\, p_1(\mathbf{y}, \mathbf{z})\, \mathrm{d}\mathbf{y}\, \mathrm{d}\mathbf{z} \\
&= \int_{\mathcal{A}\cap\mathcal{B}} \int_{\mathcal{Y}^n} \mathrm{I}\{\mathbf{y} \in \mathcal{D}_0\}\, r(\mathbf{y}, \mathbf{z})\, q_1(\mathbf{y}, \mathbf{z})\, \mathrm{d}\mathbf{y}\, \mathrm{d}\mathbf{z} \\
&\geq e^{-\alpha n} \int_{\mathcal{A}\cap\mathcal{B}} \int_{\mathcal{Y}^n} \mathrm{I}\{\mathbf{y} \in \mathcal{D}_0\}\, q_1(\mathbf{y}, \mathbf{z})\, \mathrm{d}\mathbf{y}\, \mathrm{d}\mathbf{z} \\
&= e^{-\alpha n} \int_{\mathcal{A}\cap\mathcal{B}} \int_{\mathcal{Y}^n} \mathrm{I}\{\mathbf{y} \in \mathcal{D}_0\}\, q_1(\mathbf{y}|\mathbf{z}) q_1(\mathbf{z})\, \mathrm{d}\mathbf{y}\, \mathrm{d}\mathbf{z} \\
&= e^{-\alpha n} \int_{\mathcal{A}\cap\mathcal{B}} p(\mathbf{z}) \int_{\mathcal{Y}^n} \mathrm{I}\{\mathbf{y} \in \mathcal{D}_0\}\, q_1(\mathbf{y}|\mathbf{z})\, \mathrm{d}\mathbf{y}\, \mathrm{d}\mathbf{z}. \qquad (48)
\end{aligned}$$

Similarly,

$$\mathsf{P}_0(\mathbf{Y} \in \mathcal{D}_1) \geq e^{-\alpha n} \int_{\mathcal{A}\cap\mathcal{B}} p(\mathbf{z}) \int_{\mathcal{Y}^n} \mathrm{I}\{\mathbf{y} \in \mathcal{D}_1\}\, q_0(\mathbf{y}|\mathbf{z})\, \mathrm{d}\mathbf{y}\, \mathrm{d}\mathbf{z}. \qquad (49)$$

Averaging (48) and (49) over the uniform prior of $H$ we obtain

$$\begin{aligned}
&\mathsf{P}(\text{error}) \\
&\quad = \frac{1}{2}\, \mathsf{P}_1(\mathbf{Y} \in \mathcal{D}_0) + \frac{1}{2}\, \mathsf{P}_0(\mathbf{Y} \in \mathcal{D}_1) \\
&\quad \geq e^{-\alpha n} \int_{\mathcal{A}\cap\mathcal{B}} p(\mathbf{z}) \left( \frac{1}{2} \int_{\mathcal{Y}^n} \mathrm{I}\{\mathbf{y} \in \mathcal{D}_0\} q_1(\mathbf{y}|\mathbf{z})\, \mathrm{d}\mathbf{y} + \frac{1}{2} \int_{\mathcal{Y}^n} \mathrm{I}\{\mathbf{y} \in \mathcal{D}_1\} q_0(\mathbf{y}|\mathbf{z}) \right) \mathrm{d}\mathbf{y} \\
&\quad = e^{-\alpha n} \int_{\mathcal{A}\cap\mathcal{B}} p(\mathbf{z})\, \mathsf{Q}(\text{error}|\mathbf{z})\, \mathrm{d}\mathbf{z} \\
&\quad \geq e^{-\alpha n}\, \mathcal{Q}\!\left( \sqrt{\frac{n\mathsf{P}}{(1-2\epsilon)\sigma^2}} \right) \int_{\mathcal{A}\cap\mathcal{B}} p(\mathbf{z})\, \mathrm{d}\mathbf{z} \\
&\quad \geq \epsilon\, e^{-\alpha n}\, \mathcal{Q}\!\left( \sqrt{\frac{n\mathsf{P}}{(1-2\epsilon)\sigma^2}} \right), \qquad (50)
\end{aligned}$$

where the first line follows from our assumption that the prior is uniform; the second from (48) and (49); the third because under $\mathsf{Q}$ the hypothesis $H$ is independent of $\mathbf{Z}$; the fourth from (24) and because for $\mathbf{z} \in \mathcal{A}$ the transmitted energy is bounded by $n\mathsf{P}/(1-2\epsilon)$ (44); and the fifth line from (46).

It follows from (50) and from the definition of $\alpha$ (47b) that

$$\overline{\lim_{n\to\infty}} -\frac{1}{n} \log p_{\mathrm{e}}^{\mathrm{a.s.}}\bigl(\mathsf{P}/\sigma^2, \mathsf{P}_{\mathrm{FB}}/\sigma_{\mathrm{FB}}^2, n\bigr) \leq \frac{2\sqrt{\beta \mathsf{P}_{\mathrm{FB}}}}{\sigma_{\mathrm{FB}}^2} + \frac{1}{2}\frac{\mathsf{P}}{(1-2\epsilon)\sigma^2}.$$

Since $\epsilon$ and $\delta$ can be taken as small as we wish, and since by (40)

$$\lim_{\delta\downarrow 0} \beta(\delta) = \mathsf{P}_{\mathrm{FB}} + \sigma_{\mathrm{FB}}^2,$$

we conclude that

$$\overline{\lim_{n\to\infty}} -\frac{1}{n} \log p_{\mathrm{e}}^{\mathrm{a.s.}}\bigl(\mathsf{P}/\sigma^2, \mathsf{P}_{\mathrm{FB}}/\sigma_{\mathrm{FB}}^2, n\bigr) \leq \frac{2\sqrt{(\mathsf{P}_{\mathrm{FB}} + \sigma_{\mathrm{FB}}^2)\mathsf{P}_{\mathrm{FB}}}}{\sigma_{\mathrm{FB}}^2} + \frac{1}{2}\frac{\mathsf{P}}{\sigma^2},$$

thus establishing (11).

**Note:** The lower bound on the error probability in the case of almost-sure block power constraints easily extends to the case of many codewords. In this case one typically denotes the message by $M$ instead of $H$. The analysis goes through with hardly any change to show that at any rate $R$, the error exponent $E_{\mathrm{FB,a.s.}}(R)$ with an almost-sure block power constraint is upper-bounded by

$$E_{\mathrm{FB,a.s.}}(R) \leq E_{\mathrm{NoFB}}(R) + \frac{2\sqrt{(\mathsf{P}_{\mathrm{FB}} + \sigma_{\mathrm{FB}}^2)\mathsf{P}_{\mathrm{FB}}}}{\sigma_{\mathrm{FB}}^2}, \tag{51}$$

where $E_{\mathrm{NoFB}}(R)$ is the reliability function of the forward channel in the absence of feedback.

# 5 Achievability under A.S. Power Constraints

We next describe a sequence of codes that proves (12).

## 5.1 The Scheme

Fix some $0 < \delta < 1$. At time-1 we send the message using antipodal signaling:

$$X_1 = \begin{cases} +\sqrt{n-1}\sqrt{\mathsf{P}} & \text{if } H = 0, \\ -\sqrt{n-1}\sqrt{\mathsf{P}} & \text{if } H = 1. \end{cases} \tag{52}$$

At the subsequent $(n-2)$ times $k = 2, 3, \ldots, n-1$ the transmitter is silent

$$X_k = 0, \quad k = 2, 3, \ldots, n-1. \tag{53}$$

Based on the sign of $Y_1$, the receiver forms a *tentative decision* on $H$ and feeds it back via the feedback link using antipodal signaling:

$$U_1 = U_2 = \cdots = U_{n-1} = \begin{cases} +\sqrt{\mathsf{P}_{\mathrm{FB}}} & \text{if } Y_1 > 0, \\ -\sqrt{\mathsf{P}_{\mathrm{FB}}} & \text{if } Y_1 < 0. \end{cases} \tag{54}$$

The encoder chooses the time-$n$ transmitted symbol $X_n$ based on $H$ and $Z_1, \ldots Z_{n-1}$ as follows. If

$$\frac{1}{n-1} \sum_{k=1}^{n-1} Z_k > +\sqrt{\mathsf{P}_{\mathrm{FB}}}(1-\delta) \;\text{ and }\; H = 0, \tag{55}$$

then the transmitter assumes that the tentative decision that the receiver made based on $Y_1$ is correct and it sends nothing at time $n$, i.e., it sets $X_n = 0$. Likewise, if

$$\frac{1}{n-1} \sum_{k=1}^{n-1} Z_k < -\sqrt{\mathsf{P}_{\mathrm{FB}}}(1-\delta) \;\text{ and }\; H = 1, \tag{56}$$

then the transmitter sets $X_n = 0$. In all other cases we say that a *re-transmission* event Re-Tx occurred and the transmitter retransmits the message using antipodal signaling with a huge instantaneous power, i.e., by sending $X_n = \pm\sqrt{\mathsf{P}/\gamma}$, where $\gamma$ is the probability of retransmission (which we shall soon see is exponentially small). Thus

$$X_n = \begin{cases} +\sqrt{\mathsf{P}/\gamma} & \text{if } (n-1)^{-1} \sum_{k=1}^{n-1} Z_k < \sqrt{\mathsf{P}_{\mathrm{FB}}}(1-\delta) \text{ and } H = 0, \\ -\sqrt{\mathsf{P}/\gamma} & \text{if } (n-1)^{-1} \sum_{k=1}^{n-1} Z_k > -\sqrt{\mathsf{P}_{\mathrm{FB}}}(1-\delta) \text{ and } H = 1, \\ 0 & \text{otherwise.} \end{cases} \tag{57}$$

To make its *final decision*, the receiver considers $Y_n$ and compares it to a threshold $\Upsilon$. (We shall later choose $\Upsilon$ to equal the blocklength $n$.) If $Y_n > \Upsilon$, then the receiver assumes that a retransmission took place and decides "$H = 0$." Similarly, if $Y_n < -\Upsilon$, it assumes that a retransmission occurred and decides "$H = 1$." Otherwise, if $|Y_n| \leq \Upsilon$, it decides that a retransmission did not take place and that its tentative decision was correct: it declares "$H = 0$" if $Y_1 > 0$ and declares "$H = 1$" if $Y_1 < 0$. If $\hat{H}$ denotes the final decision, then

$$\hat{H} = \begin{cases} 0 & \text{if } Y_n > \Upsilon \text{ or if } \Bigl(|Y_n| < \Upsilon \text{ and } Y_1 > 0\Bigr), \\ 1 & \text{if } Y_n < -\Upsilon \text{ or if } \Bigl(|Y_n| < \Upsilon \text{ and } Y_1 < 0\Bigr). \end{cases} \tag{58}$$

## 5.2 Analysis

It is straightforward to see that the expected average transmitted power on the forward link is $\mathsf{P}$ and that with probability one the average transmitted power on the feedback link does not exceed $\mathsf{P}_{\mathrm{FB}}$. We focus on the probability of error. We shall assume throughout that $H = 0$; the analysis of the case where $H = 1$ is very similar. When $H = 0$, a retransmission occurs if the event

$$\mathcal{G} \triangleq \left\{ \frac{1}{n-1} \sum_{k=1}^{n-1} Z_k \leq \sqrt{\mathsf{P}}(1-\delta) \right\} \tag{59}$$

occurs. Substituting the event $\{Y_1 > 0\}$ for $\mathcal{A}$ and $\mathcal{G}$ for $\mathcal{B}$ in the inequality

$$\Pr(\mathcal{A} \cup \mathcal{B}) \leq \Pr(\mathcal{B} \,|\, \mathcal{A}) + \Pr(\mathcal{A}^{\mathrm{c}}), \quad \Pr(\mathcal{A}) > 0 \tag{60}$$

yields that the conditional probability of a retransmission $\mathsf{P}_0(\text{Re-Tx})$ is upper-bounded by

$$\begin{aligned} \mathsf{P}_0(\text{Re-Tx}) &\leq \Pr\left[ \frac{1}{n-1} \sum_{k=1}^{n-1} Z_k \leq \sqrt{\mathsf{P}}(1-\delta) \,\middle|\, Y_1 > 0 \right] + \mathsf{P}_0\bigl(Y_1 \leq 0\bigr) & (61) \\ &= \mathcal{Q}\left( \frac{\sqrt{n-1}\sqrt{\mathsf{P}_{\mathrm{FB}}}\delta}{\sigma_{\mathrm{FB}}} \right) + \mathcal{Q}\left( \frac{\sqrt{n-1}\sqrt{\mathsf{P}}}{\sigma} \right). & (62) \end{aligned}$$

By a similar argument, one can show that this also upper-bounds the probability of retransmission conditional on $H = 1$. Thus,

$$\gamma = \mathsf{P}(\text{Re-Tx}) \leq \mathcal{Q}\left(\frac{\sqrt{n-1}\sqrt{\mathsf{P}}}{\sigma}\right) + \mathcal{Q}\left(\frac{\sqrt{n-1}\sqrt{\mathsf{P}_{\text{FB}}}\delta}{\sigma_{\text{FB}}}\right). \tag{63}$$

Notice that for a fixed $\delta > 0$, it follows from (63) that $\gamma$ tends to zero exponentially in $n$. Consequently the amplitude $\sqrt{\mathsf{P}/\gamma}$ in the retransmission phase is much larger than the noise variance, and, in fact, the ratio

$$\frac{\sqrt{\mathsf{P}/\gamma}}{\sigma} \tag{64}$$

grows exponentially in $n$. If we now set $\Upsilon = n$, then the probabilities

$$\Pr\bigl[|Y_n| > \Upsilon \bigm| X_n = 0\bigr] \tag{65a}$$

$$\Pr\bigl[Y_n < \Upsilon \bigm| X_n = \sqrt{\mathsf{P}/\gamma}\,\bigr] \tag{65b}$$

and

$$\Pr\bigl[Y_n > -\Upsilon \bigm| X_n = -\sqrt{\mathsf{P}/\gamma}\,\bigr] \tag{65c}$$

will all decay faster than exponentially to zero. We conclude that the probability that a retransmission takes place and a decoding error occurs decays faster than exponentially in $n$. Thus,

$$\liminf -\frac{1}{n}\log \Pr[\text{Re-Tx and error}] = +\infty. \tag{66}$$

The dominant error that will determine the exponential decay of the probability of error of our scheme is thus the probability that a retransmission does not take place and an error occurs. As before, this event does not depend on the hypothesis by symmetry. If $H = 0$, then the event that no retransmission takes place and yet an error occurs is equivalent to the tentative decision being wrong

$$\bigl\{Y_n < 0\bigr\} \tag{67a}$$

and the transmitter not realizing this

$$\left\{\frac{1}{n-1}\sum_{k=1}^{n-1} Z_k > \sqrt{\mathsf{P}_{\text{FB}}}(1-\delta)\right\}. \tag{67b}$$

The latter two events are independent and thus

$$\begin{aligned}\mathsf{P}_0(\text{no Re-Tx and error}) &= \mathsf{P}_0\Big((n-1)^{-1}\sum_{k=2}^{n} Z_k > \sqrt{\mathsf{P}_{\text{FB}}}(1-\delta)\Big)\mathsf{P}_0(Y_1<0)\\ &= \mathcal{Q}\left(\frac{\sqrt{n-1}\mathsf{P}_{\text{FB}}(2-\delta)}{\sigma_{\text{FB}}}\right)\mathcal{Q}\left(\frac{\sqrt{n-1}\mathsf{P}}{\sigma}\right). \qquad (68)\end{aligned}$$

An analogous expression can be derived conditional on $H=1$ to conclude that

$$\mathsf{P}(\text{no Re-Tx and error}) = \mathcal{Q}\left(\frac{\sqrt{n-1}\sqrt{\mathsf{P}_{\text{FB}}}(2-\delta)}{\sigma_{\text{FB}}}\right)\mathcal{Q}\left(\frac{\sqrt{n-1}\sqrt{\mathsf{P}}}{\sigma}\right). \quad (69)$$

Combining (69) with (66) demonstrates that our scheme achieves an error exponent of

$$\frac{1}{2}\frac{\mathsf{P}}{\sigma^2}+\frac{1}{2}\frac{\mathsf{P}_{\text{FB}}(2-\delta)^2}{\sigma^2_{\text{FB}}}. \qquad (70)$$

By considering the limit of $\delta \downarrow 0$ we obtain that any exponent smaller than

$$\frac{\mathsf{P}}{2\sigma^2}+\frac{2\mathsf{P}_{\text{FB}}}{\sigma^2_{\text{FB}}} \qquad (71)$$

is achievable. In fact, by letting $\delta$ tends to zero very slowly with $n$, we can achieve the exponent (71) and thus establish (12).

# 6 Inachievability under Expected Power Constraints

In this section we prove (14). We begin with some definitions.

## 6.1 Some Definitions

Given $\alpha_1, \alpha_2 > 0$ and some blocklength-$n$ coding scheme, we define the set

$$\mathcal{T}_y = \Big\{\mathbf{y}\in\mathcal{Y}^n : \|\mathbf{y}\|^2 < n\alpha_1^2,\ \sum_{k=1}^{n} g_k^2\big(y^k\big) < n\alpha_2^2\Big\}. \qquad (72)$$

Notice that the set $\mathcal{T}_y$ depends on $\alpha_1$, $\alpha_2$, the blocklength $n$, and the feedback-channel encoding rule under consideration, but our notation does not make this explicit.

Given $\beta_1, \beta_2 > 0$ and some blocklength-$n$ coding scheme we define

$$\mathcal{T}_z^{(0)} = \Big\{ \mathbf{z} \in \mathcal{Z}^n : \|\mathbf{z}\|^2 < n\beta_1^2, \ \sum_{k=1}^{n} f_k^2\big(0, z^{k-1}\big) < n\beta_2^2 \Big\}. \tag{73}$$

This set depends on $\beta_1$, $\beta_2$, the blocklength $n$, and the forward-channel encoder under consideration.

## 6.2 Preliminary Lemmas

We present here some lemmas that will be useful in proving (14).

**Lemma 1.** *The following inequality holds:*

$$p_1(\mathbf{y}) \leq (2\pi\sigma^2)^{-n/2}, \quad \mathbf{y} \in \mathcal{Y}^n. \tag{74}$$

*Proof.* By (16)

$$\begin{aligned} p_1(\mathbf{y}, \mathbf{z}) &= \prod_{k=1}^{n} w\Big(y_k \big| f_k\big(1, z^{k-1}\big)\Big) \prod_{k=1}^{n} p_1\big(z_k \big| y^k, z^{k-1}\big) & (75) \\ &\leq (2\pi\sigma^2)^{-n/2} \prod_{k=1}^{n} p_1\big(z_k \big| y^k, z^{k-1}\big), \quad \mathbf{y} \in \mathcal{Y}^n. & (76) \end{aligned}$$

For a fixed $\mathbf{y} \in \mathcal{Y}^n$ the product

$$\prod_{k=1}^{n} p_1\big(z_k \big| y^k, z^{k-1}\big) \tag{77}$$

is a density on $\mathcal{Z}^n$. Integrating (76) over $\mathbf{z}$ we thus obtain

$$\begin{aligned} p_1(\mathbf{y}) &= \int_{\mathcal{Z}^n} p_1(\mathbf{y}, \mathbf{z}) \, \mathrm{d}\mathbf{z} \\ &\leq (2\pi\sigma^2)^{-n/2} \int_{\mathcal{Z}^n} \prod_{k=1}^{n} p_1\big(z_k \big| y^k, z^{k-1}\big) \, \mathrm{d}\mathbf{z} \\ &= (2\pi\sigma^2)^{-n/2}, \quad \mathbf{y} \in \mathcal{Y}^n. \end{aligned}$$

□

**Lemma 2.** *The following inequality holds:*

$$p_0(\mathbf{z}) \le (2\pi\sigma_{\mathrm{FB}}^2)^{-n/2}, \quad \mathbf{z} \in \mathcal{Z}^n. \tag{78}$$

*Proof.* By (16)

$$p_0(\mathbf{y},\mathbf{z}) = \prod_{k=1}^{n} p_0\Big(y_k|f_k\big(0,z^{k-1}\big)\Big) \prod_{k=1}^{n} w_{\mathrm{FB}}\Big(z_k|g_k\big(y^k\big)\Big) \tag{79}$$

$$\le (2\pi\sigma_{\mathrm{FB}}^2)^{-n/2} \prod_{k=1}^{n} p_0\Big(y_k|f_k\big(0,z^{k-1}\big)\Big), \quad \mathbf{z} \in \mathcal{Z}^n. \tag{80}$$

For a fixed $\mathbf{z} \in \mathcal{Z}^n$, the product

$$\prod_{k=1}^{n} p_0\Big(y_k|f_k\big(0,z^{k-1}\big)\Big)$$

is a density on $\mathcal{Y}^n$. Integrating over $\mathbf{y}$ we thus obtain

$$\begin{aligned} p_0(\mathbf{z}) &= \int_{\mathcal{Y}^n} p_0(\mathbf{y},\mathbf{z})\,\mathrm{d}\mathbf{y} \\ &\le (2\pi\sigma_{\mathrm{FB}}^2)^{-n/2} \int_{\mathcal{Y}^n} \prod_{k=1}^{n} p_0\Big(y_k|f_k\big(0,z^{k-1}\big)\Big)\,\mathrm{d}\mathbf{y} \\ &= (2\pi\sigma_{\mathrm{FB}}^2)^{-n/2}, \quad \mathbf{z} \in \mathcal{Z}^n. \end{aligned}$$

□

**Lemma 3.** *The following inequality holds:*

$$p_0(\mathbf{y},\mathbf{z}) \ge (4\pi^2\sigma^2\sigma_{\mathrm{FB}}^2)^{-n/2} \exp\bigg(-n\Big(\frac{(\alpha_1+\beta_2)^2}{2\sigma^2} + \frac{(\beta_1+\alpha_2)^2}{2\sigma_{\mathrm{FB}}^2}\Big)\bigg), \quad \mathbf{y} \in \mathcal{T}_y,\ \mathbf{z} \in \mathcal{T}_z^{(0)}. \tag{81}$$

*Proof.* By (16) we note that

$$p_0(\mathbf{y},\mathbf{z}) = \prod_{k=1}^{n} w\Big(y_k|f_k\big(0,z^{k-1}\big)\Big)\ \prod_{k=1}^{n} w_{\mathrm{FB}}\Big(z_k|g_k\big(y^k\big)\Big), \tag{82}$$

and we proceed to lower-bound each of the products separately for $\mathbf{y} \in \mathcal{T}_y$ and $\mathbf{z} \in \mathcal{T}_z^{(0)}$. For such $\mathbf{y}$ and $\mathbf{z}$ we have

$$\begin{aligned}\prod_{k=1}^{n} w\Big(y_k \big| f_k\big(0, z^{k-1}\big)\Big) &= (2\pi\sigma^2)^{-n/2} \exp\bigg(-\frac{1}{2\sigma^2}\sum_{k=1}^{n}\Big(y_k - f_k\big(0, z^{k-1}\big)\Big)^2\bigg) \\ &\geq (2\pi\sigma^2)^{-n/2} \exp\left(-\frac{1}{2\sigma^2}\left(\|\mathbf{y}\| + \sqrt{\sum_{k=1}^{n} f_k^2\big(0, z^{k-1}\big)}\right)^2\right) \\ &\geq (2\pi\sigma^2)^{-n/2} \exp\Big(-\frac{1}{2\sigma^2}\Big(\sqrt{n}\alpha_1 + \sqrt{n}\beta_2\Big)^2\Big) \\ &= (2\pi\sigma^2)^{-n/2} \exp\Big(-n\frac{(\alpha_1+\beta_2)^2}{2\sigma^2}\Big). \end{aligned} \tag{83}$$

We next turn to the second product. For $\mathbf{y}$ and $\mathbf{z}$ as above

$$\begin{aligned}\prod_{k=1}^{n} w_{\mathrm{FB}}\Big(z_k \big| g_k\big(y^k\big)\Big) &= (2\pi\sigma_{\mathrm{FB}}^2)^{-n/2} \exp\bigg(-\frac{1}{2\sigma_{\mathrm{FB}}^2}\sum_{k=1}^{n}\Big(z_k - g_k\big(y^k\big)\Big)^2\bigg) \\ &\geq (2\pi\sigma_{\mathrm{FB}}^2)^{-n/2} \exp\left(-\frac{1}{2\sigma_{\mathrm{FB}}^2}\left(\|\mathbf{z}\| + \sqrt{\sum_{k=1}^{n} g_k^2\big(y^k\big)}\right)^2\right) \\ &\geq (2\pi\sigma_{\mathrm{FB}}^2)^{-n/2} \exp\Big(-\frac{1}{2\sigma_{\mathrm{FB}}^2}\Big(\sqrt{n}\beta_1 + \sqrt{n}\alpha_2\Big)^2\Big) \\ &= (2\pi\sigma_{\mathrm{FB}}^2)^{-n/2} \exp\Big(-n\frac{(\beta_1+\alpha_2)^2}{2\sigma_{\mathrm{FB}}^2}\Big). \end{aligned} \tag{84}$$

Inequalities (84) and (83) combine with (82) to prove the lemma. □

**Lemma 4.** *The following inequality holds:*

$$p_0(\mathbf{y}) \geq (2\pi\sigma^2)^{-n/2} \exp\Big(-n\Big(\frac{(\alpha_1+\beta_2)^2}{2\sigma^2} + \frac{(\beta_1+\alpha_2)^2}{2\sigma_{\mathrm{FB}}^2}\Big)\Big)\mathsf{P}_0\big(\mathbf{Z} \in \mathcal{T}_z^{(0)}\big), \quad \mathbf{y} \in \mathcal{T}_y. \tag{85}$$

*Proof.* For every $\mathbf{y} \in \mathcal{T}_y$ we have

$$\begin{aligned} p_0(\mathbf{y}) &= \int_{\mathcal{Z}^n} p_0(\mathbf{y}, \mathbf{z}) \, \mathrm{d}\mathbf{z} \\ &\geq \int_{\mathcal{T}_z^{(0)}} p_0(\mathbf{y}, \mathbf{z}) \, \mathrm{d}\mathbf{z} \\ &\geq (4\pi^2\sigma^2\sigma_{\mathrm{FB}}^2)^{-n/2} \exp\biggl(-n\Bigl(\frac{(\alpha_1+\beta_2)^2}{2\sigma^2} + \frac{(\beta_1+\alpha_2)^2}{2\sigma_{\mathrm{FB}}^2}\Bigr)\biggr) \int_{\mathcal{T}_z^{(0)}} \mathrm{d}\mathbf{z}, \end{aligned}$$

where the last inequality follows from Lemma 3, i.e., from (81). It remains to lower-bound the volume of $\mathcal{T}_z^{(0)}$ as follows:

$$\begin{aligned} \int_{\mathcal{T}_z^{(0)}} \mathrm{d}\mathbf{z} &\geq \frac{\mathsf{P}_0\bigl(\mathbf{Z} \in \mathcal{T}_z^{(0)}\bigr)}{\sup_{\mathbf{z}\in\mathcal{T}_z^{(0)}} p_0(\mathbf{z})} \\ &\geq \frac{\mathsf{P}_0\bigl(\mathbf{Z} \in \mathcal{T}_z^{(0)}\bigr)}{(2\pi\sigma_{\mathrm{FB}}^2)^{-n/2}}, \end{aligned}$$

where the last inequality follows from Lemma 2, i.e., from (78). □

**Lemma 5.** *The following inequality holds:*

$$\frac{p_0(\mathbf{y})}{p_1(\mathbf{y})} \geq \exp\biggl(-n\Bigl(\frac{(\alpha_1+\beta_2)^2}{2\sigma^2} + \frac{(\beta_1+\alpha_2)^2}{2\sigma_{\mathrm{FB}}^2}\Bigr)\biggr)\mathsf{P}_0\bigl(\mathbf{Z} \in \mathcal{T}_z^{(0)}\bigr), \quad \mathbf{y} \in \mathcal{T}_y. \tag{86}$$

*Proof.* Follows from Lemma 4 and Lemma 1, i.e., from (85) and (74). □

**Lemma 6.** *The following inequality holds:*

$$\begin{aligned} &\mathsf{P}_0\bigl(\mathbf{Y} \in \mathcal{T}_y \cap \mathcal{D}_1\bigr) \geq \\ &\exp\biggl(-n\Bigl(\frac{(\alpha_1+\beta_2)^2}{2\sigma^2} + \frac{(\beta_1+\alpha_2)^2}{2\sigma_{\mathrm{FB}}^2}\Bigr)\biggr)\mathsf{P}_0\bigl(\mathbf{Z} \in \mathcal{T}_z^{(0)}\bigr)\mathsf{P}_1\bigl(\mathbf{Y} \in \mathcal{T}_y \cap \mathcal{D}_1\bigr). \end{aligned} \tag{87}$$

*Proof.* This follows from Lemma 6, i.e., from (87) as follows:

$$\begin{aligned} &\mathsf{P}_0\bigl(\mathbf{Y} \in \mathcal{T}_y \cap \mathcal{D}_1\bigr) \\ &\quad = \int_{\mathcal{T}_y\cap\mathcal{D}_1} \frac{p_0(\mathbf{y})}{p_1(\mathbf{y})} p_1(\mathbf{y}) \, \mathrm{d}\mathbf{y} \\ &\quad \geq \exp\biggl(-n\Bigl(\frac{(\alpha_1+\beta_2)^2}{2\sigma^2} + \frac{(\beta_1+\alpha_2)^2}{2\sigma_{\mathrm{FB}}^2}\Bigr)\biggr)\mathsf{P}_0\bigl(\mathbf{Z} \in \mathcal{T}_z^{(0)}\bigr) \int_{\mathcal{T}_y\cap\mathcal{D}_1} p_1(\mathbf{y}) \, \mathrm{d}\mathbf{y} \\ &\quad = \exp\biggl(-n\Bigl(\frac{(\alpha_1+\beta_2)^2}{2\sigma^2} + \frac{(\beta_1+\alpha_2)^2}{2\sigma_{\mathrm{FB}}^2}\Bigr)\biggr)\mathsf{P}_0\bigl(\mathbf{Z} \in \mathcal{T}_z^{(0)}\bigr)\mathsf{P}_1\bigl(\mathbf{Y} \in \mathcal{T}_y \cap \mathcal{D}_1\bigr). \end{aligned}$$ □

### 6.3 A Proof of (14)

We now use Lemma 6 to prove (14). Suppose we are given a sequence of codes, with a code corresponding to each blocklength $n$. To establish (14) we need to show that from every subsequence $\{n_k\}$ of blocklengths tending to infinity we can extract a subsequence $\{n'_k\}$ such that

$$\overline{\lim_{k\to\infty}} -\frac{1}{n'_k}\log \mathsf{P}(\text{error}, n'_k) \le \frac{\left(\sqrt{\mathsf{P}+\sigma^2}+\sqrt{\mathsf{P}}\right)^2}{\sigma^2} + \frac{\left(\sqrt{\mathsf{P}_{\text{FB}}+\sigma^2_{\text{FB}}}+\sqrt{\mathsf{P}_{\text{FB}}}\right)^2}{\sigma^2_{\text{FB}}}, \tag{88}$$

where $\mathsf{P}(\text{error}, n'_k)$ denotes the probability of error of the code corresponding to the blocklength $n'_k$. We refer to this as "our claim" and proceed to prove it. We thus start with some subsequence $\{n_k\}$ and proceed to prove the existence of a subsequence $\{n'_k\}$ tending to infinity for which (88) holds.

Before proving our claim we make several reductions. We first assume that the subsequence $\{n_k\}$ is such that

$$\lim_{n_k\to\infty} \mathsf{P}_\nu\big(\mathbf{Y}\in\mathcal{D}_\nu\big) = 1, \quad \nu = 0, 1. \tag{89}$$

There is no loss in generality in making this assumption because otherwise we can pick $\{n'_k\}$ to be a subsequence for which for some $\nu \in \{0, 1\}$ the limit as $n'_k \to \infty$ of $\mathsf{P}_\nu(\mathbf{Y}\in\mathcal{D}_\nu)$ is not 1, in which case the LHS of (88) zero and hence trivially smaller than its RHS.

Before stating the second reduction we introduce some notation. For a given sequence of codes indexed by the blocklength-$n$, we define for every blocklength $n$

$$\mathsf{P}_n^{(\nu)} = \frac{1}{n}\mathsf{E}\left[\sum_{k=1}^{n} f_k^2\big(\nu, Z^{k-1}\big)\,\middle|\, H=\nu\right], \quad \nu = 0, 1, \tag{90}$$

where $f_1, \ldots, f_n$ are the mappings that define the forward-channel encoder of the code indexed by $n$. (The dependence of the encoding mappings on the blocklength $n$ is omitted for brevity.) Note that by the block power constraint (8)

$$\mathsf{P}_n^{(0)} + \mathsf{P}_n^{(1)} \le 2\mathsf{P}, \quad n = 1, 2, \ldots \tag{91}$$

The second reduction we make is in assuming that the subsequence $\{n_k\}$ is such that the limits defining $\mathsf{P}^{(0)}$ and $\mathsf{P}^{(1)}$ as

$$\mathsf{P}^{(\nu)} = \lim_{k\to\infty} \mathsf{P}_{n_k}^{(\nu)}, \quad \nu = 0, 1 \tag{92}$$

exist and satisfy

$$\mathsf{P}^{(0)} + \mathsf{P}^{(1)} \le 2\mathsf{P}. \tag{93}$$

There is no loss in generality in making this assumption because the more general case where the above does not necessarily hold follows from the less-general case by applying the less-general result to the subsequence of $\{n_k\}$ which we denote by $\{\tilde{n}'_k\}$ and which we define as follows. We pick a subsequence $\{\tilde{n}_k\}$ of $\{n_k\}$ for which the limit of $\mathsf{P}^{(0)}_{\tilde{n}_k}$ exists; we call the limit $\mathsf{P}^{(0)}$; we define the subsequence $\{\tilde{n}'_k\}$ of $\{\tilde{n}_k\}$—and hence also of $\{n_k\}$—as a subsequence of $\{\tilde{n}_k\}$ for which the limit of $\mathsf{P}^{(1)}_{\tilde{n}'_k}$ also exists; and we finally define $\mathsf{P}^{(1)}$ as this limit. We then observe that (93) then follows from (91).

Finally, we define for every blocklength $n$

$$\mathsf{P}^{(\nu)}_{\mathrm{FB,n}} \triangleq \frac{1}{n}\mathsf{E}\left[\sum_{k=1}^{n} g_k^2\bigl(Y^k\bigr)\,\middle|\, H=\nu\right], \quad \nu = 0,1, \tag{94}$$

where $g_1, \ldots, g_n$ are the mappings that define the feedback-channel encoder. We now assume that the limits defining $\mathsf{P}^{(0)}_{\mathrm{FB}}$ and $\mathsf{P}^{(1)}_{\mathrm{FB}}$ as

$$\mathsf{P}^{(\nu)}_{\mathrm{FB}} = \lim_{k\to\infty} \mathsf{P}^{(\nu)}_{\mathrm{FB},n_k}, \quad \nu = 0,1 \tag{95}$$

exist and satisfy

$$\mathsf{P}^{(0)}_{\mathrm{FB}} + \mathsf{P}^{(1)}_{\mathrm{FB}} \le \mathsf{P}_{\mathrm{FB}}. \tag{96}$$

This entails no loss of generality by arguments analogous to those we used for the second reduction.

We are now ready to prove our claim subject to the above assumptions. Let $\epsilon > 0$ (smaller than $1/2$) be fixed. Choose

$$\alpha_1^2 = a_1^2\bigl(\mathsf{P}^{(1)} + \sigma^2\bigr)(1+\epsilon) \tag{97}$$

$$\alpha_2^2 = a_2^2\mathsf{P}^{(1)}_{\mathrm{FB}}(1+\epsilon) \tag{98}$$

$$\beta_1^2 = b_1^2\bigl(\mathsf{P}^{(0)}_{\mathrm{FB}} + \sigma^2_{\mathrm{FB}}\bigr)(1+\epsilon) \tag{99}$$

$$\beta_2^2 = b_2^2\mathsf{P}^{(0)}(1+\epsilon), \tag{100}$$

where the positive numbers $a_1$, $a_2$, $b_1$, and $b_2$ satisfy

$$\frac{1}{a_1^2} + \frac{1}{a_2^2} \le 1 \tag{101}$$

and

$$\frac{1}{b_1^2} + \frac{1}{b_2^2} \le 1. \tag{102}$$

Let the sets $\mathcal{T}_y$ and $\mathcal{T}_z^{(0)}$ be as defined in (72) and (73). Using the bound

$$\Pr\big(\mathcal{A} \cap \mathcal{B}\big) = \Pr\big(\mathcal{A}\big) + \Pr\big(\mathcal{B}\big) - \Pr\big(\mathcal{A} \cup \mathcal{B}\big) \ge \Pr\big(\mathcal{A}\big) + \Pr\big(\mathcal{B}\big) - 1$$

and Markov's inequality we obtain

$$\begin{aligned}
\mathsf{P}_1\big(\mathbf{Y} \in \mathcal{T}_y\big) &= \mathsf{P}_1\Big(\|\mathbf{Y}\|^2 < n\alpha_1^2,\ \sum_{k=1}^{n} g_k^2(Y^k) < n\alpha_2^2\Big) \\
&\ge \mathsf{P}_1\Big(\|\mathbf{Y}\|^2 < n\alpha_1^2\Big) + \mathsf{P}_1\Big(\sum_{k=1}^{n} g_k^2(Y^k) < n\alpha_2^2\Big) - 1 \\
&\ge 1 - \frac{\mathsf{E}\big[\|\mathbf{Y}\|^2 \,\big|\, H = 1\big]}{n\alpha_1^2} + 1 - \frac{\mathsf{E}\Big[\sum_{k=1}^{n} g_k^2(Y^k) \,\Big|\, H = 1\Big]}{n\alpha_2^2} - 1 \\
&= 1 - \left(\frac{\mathsf{P}_n^{(1)} + \sigma^2}{\alpha_1^2} + \frac{\mathsf{P}_{\text{FB,n}}^{(1)}}{\alpha_2^2}\right), \quad n = 1, 2, \ldots
\end{aligned}$$

It follows from (90), (95), and (101) that for all sufficiently large $n_k$

$$\mathsf{P}_1\big(\mathbf{Y} \in \mathcal{T}_y\big) > \frac{\epsilon}{2}. \tag{103}$$

From this and (89) we can infer that for all sufficiently large $n_k$

$$\mathsf{P}_1\big(\mathbf{Y} \in \mathcal{T}_y \cap \mathcal{D}_1\big) > \frac{\epsilon}{3}. \tag{104}$$

By similar arguments we can show that for all sufficiently large $n_k$

$$\mathsf{P}_1\big(\mathbf{Y} \in \mathcal{T}_z^{(0)}\big) > \frac{\epsilon}{2}. \tag{105}$$

From Lemma 6 we then have

$$\mathsf{P}_0(\text{error}) = \mathsf{P}_0\big(\mathbf{Y} \in \mathcal{D}_1\big) \tag{106}$$

$$\ge \mathsf{P}_0\big(\mathbf{Y} \in \mathcal{T}_y \cap \mathcal{D}_1\big) \tag{107}$$

$$\ge \frac{1}{6}\epsilon^2 \exp\bigg(-n\Big(\frac{(\alpha_1 + \beta_2)^2}{2\sigma^2} + \frac{(\beta_1 + \alpha_2)^2}{2\sigma_{\text{FB}}^2}\Big)\bigg). \tag{108}$$

Since $\epsilon$ can be chosen as small as we wish, we conclude that no error exponent higher than

$$E_1 \triangleq \inf_{a_1,a_2,b_1,b_2} \left\{ \frac{\left(a_1\sqrt{\mathsf{P}^{(1)}+\sigma^2}+b_2\sqrt{\mathsf{P}^{(0)}}\right)^2}{2\sigma^2} + \frac{\left(b_1\sqrt{\mathsf{P}_{\mathrm{FB}}^{(0)}+\sigma_{\mathrm{FB}}^2}+a_2\sqrt{\mathsf{P}_{\mathrm{FB}}^{(1)}}\right)^2}{2\sigma_{\mathrm{FB}}^2} \right\}$$

is achievable, where the infimum is over all positive $a_1, a_2, b_1, b_2$ satisfying (101) and (102).

We could, of course, have also reversed the roles of $H = 0$ and $H = 1$ to obtain that no error exponent higher than

$$E_0 \triangleq \inf_{a_1,a_2,b_1,b_2} \left\{ \frac{\left(a_1\sqrt{\mathsf{P}^{(0)}+\sigma^2}+b_2\sqrt{\mathsf{P}^{(1)}}\right)^2}{2\sigma^2} + \frac{\left(b_1\sqrt{\mathsf{P}_{\mathrm{FB}}^{(1)}+\sigma_{\mathrm{FB}}^2}+a_2\sqrt{\mathsf{P}_{\mathrm{FB}}^{(0)}}\right)^2}{2\sigma_{\mathrm{FB}}^2} \right\}$$

is achievable. Thus, our best bound is

$$\sup_{\mathsf{P}^{(0)},\mathsf{P}^{(1)},\mathsf{P}_{\mathrm{FB}}^{(0)},\mathsf{P}_{\mathrm{FB}}^{(1)}} \min\{E_1, E_0\}, \tag{109}$$

where the supremum is over all positive $\mathsf{P}^{(0)}$, $\mathsf{P}^{(1)}$, $\mathsf{P}_{\mathrm{FB}}^{(0)}$, $\mathsf{P}_{\mathrm{FB}}^{(1)}$ satisfying (93) and (96).

A suboptimal choice for $a_1, a_2, b_1, b_2$ is

$$a_1 = a_2 = b_1 = b_2 = \sqrt{2} \tag{110}$$

which leads to

$$E_1 \le E_1^{\mathrm{sub}} \triangleq \frac{\left(\sqrt{\mathsf{P}^{(1)}+\sigma^2}+\sqrt{\mathsf{P}^{(0)}}\right)^2}{\sigma^2} + \frac{\left(\sqrt{\mathsf{P}_{\mathrm{FB}}^{(0)}+\sigma_{\mathrm{FB}}^2}+\sqrt{\mathsf{P}_{\mathrm{FB}}^{(1)}}\right)^2}{\sigma_{\mathrm{FB}}^2}, \tag{111}$$

$$E_0 \le E_0^{\mathrm{sub}} \triangleq \frac{\left(\sqrt{\mathsf{P}^{(0)}+\sigma^2}+\sqrt{\mathsf{P}^{(1)}}\right)^2}{\sigma^2} + \frac{\left(\sqrt{\mathsf{P}_{\mathrm{FB}}^{(1)}+\sigma_{\mathrm{FB}}^2}+\sqrt{\mathsf{P}_{\mathrm{FB}}^{(0)}}\right)^2}{\sigma_{\mathrm{FB}}^2}. \tag{112}$$

Opening the square and using (93) and (96) we obtain

$$\frac{E_1^{\mathrm{sub}}}{2} = 1 + \frac{\mathsf{P}}{\sigma^2} + \frac{\mathsf{P}_{\mathrm{FB}}}{\sigma_{\mathrm{FB}}^2} + \frac{\sqrt{(\mathsf{P}^{(1)}+\sigma^2)\mathsf{P}^{(0)}}}{\sigma^2} + \frac{\sqrt{(\mathsf{P}_{\mathrm{FB}}^{(0)}+\sigma_{\mathrm{FB}}^2)\mathsf{P}_{\mathrm{FB}}^{(1)}}}{\sigma_{\mathrm{FB}}^2}. \tag{113}$$

and

$$\frac{E_0^{\mathrm{sub}}}{2} = 1 + \frac{\mathsf{P}}{\sigma^2} + \frac{\mathsf{P}_{\mathrm{FB}}}{\sigma_{\mathrm{FB}}^2} + \frac{\sqrt{(\mathsf{P}^{(0)} + \sigma^2)\mathsf{P}^{(1)}}}{\sigma^2} + \frac{\sqrt{(\mathsf{P}_{\mathrm{FB}}^{(1)} + \sigma_{\mathrm{FB}}^2)\mathsf{P}_{\mathrm{FB}}^{(0)}}}{\sigma_{\mathrm{FB}}^2}, \tag{114}$$

and the error exponent is upper-bounded by

$$E^{\mathrm{sub}} \triangleq \min\left\{E_1^{\mathrm{sub}}, E_0^{\mathrm{sub}}\right\}. \tag{115}$$

Now it is easy to see that the bound is maximized with $\mathsf{P}^{(0)} = \mathsf{P}^{(1)} = \mathsf{P}$ and $\mathsf{P}_{\mathrm{FB}}^{(0)} = \mathsf{P}_{\mathrm{FB}}^{(1)} = \mathsf{P}_{\mathrm{FB}}$. Indeed, the error exponent is upper-bounded by

$$\begin{aligned}
\frac{1}{2}E^{\mathrm{sub}} &\le \frac{1}{2}(E_1^{\mathrm{sub}}/2 + E_0^{\mathrm{sub}}/2) && (116)\\
&= 1 + \frac{\mathsf{P}}{\sigma^2} + \frac{\mathsf{P}_{\mathrm{FB}}}{\sigma_{\mathrm{FB}}^2} + \frac{1}{2}\frac{\sqrt{(\mathsf{P}^{(1)} + \sigma^2)\mathsf{P}^{(0)}}}{\sigma^2} + \frac{1}{2}\frac{\sqrt{(\mathsf{P}^{(0)} + \sigma^2)\mathsf{P}^{(1)}}}{\sigma^2} && (117)\\
&\quad + \frac{1}{2}\frac{\sqrt{(\mathsf{P}_{\mathrm{FB}}^{(0)} + \sigma_{\mathrm{FB}}^2)\mathsf{P}_{\mathrm{FB}}^{(1)}}}{\sigma_{\mathrm{FB}}^2} + \frac{1}{2}\frac{\sqrt{(\mathsf{P}_{\mathrm{FB}}^{(1)} + \sigma_{\mathrm{FB}}^2)\mathsf{P}_{\mathrm{FB}}^{(0)}}}{\sigma_{\mathrm{FB}}^2} && (118)\\
&\le 1 + \frac{\mathsf{P}}{\sigma^2} + \frac{\mathsf{P}_{\mathrm{FB}}}{\sigma_{\mathrm{FB}}^2} + \frac{\sqrt{\frac{1}{2}(\mathsf{P}^{(1)} + \sigma^2)\mathsf{P}^{(0)} + \frac{1}{2}(\mathsf{P}^{(0)} + \sigma^2)\mathsf{P}^{(1)}}}{\sigma^2} && (119)\\
&\quad + \frac{\sqrt{\frac{1}{2}(\mathsf{P}_{\mathrm{FB}}^{(0)} + \sigma_{\mathrm{FB}}^2)\mathsf{P}_{\mathrm{FB}}^{(1)} + \frac{1}{2}(\mathsf{P}_{\mathrm{FB}}^{(1)} + \sigma_{\mathrm{FB}}^2)\mathsf{P}_{\mathrm{FB}}^{(0)}}}{\sigma_{\mathrm{FB}}^2} && (120)\\
&= 1 + \frac{\mathsf{P}}{\sigma^2} + \frac{\mathsf{P}_{\mathrm{FB}}}{\sigma_{\mathrm{FB}}^2} + \frac{\sqrt{\sigma^2\mathsf{P} + \mathsf{P}^{(0)}\mathsf{P}^{(1)}}}{\sigma^2} + \frac{\sqrt{\sigma^2\mathsf{P}_{\mathrm{FB}} + \mathsf{P}_{\mathrm{FB}}^{(0)}\mathsf{P}_{\mathrm{FB}}^{(1)}}}{\sigma_{\mathrm{FB}}^2} && (121)\\
&\le 1 + \frac{\mathsf{P}}{\sigma^2} + \frac{\mathsf{P}_{\mathrm{FB}}}{\sigma_{\mathrm{FB}}^2} + \frac{\sqrt{\sigma^2\mathsf{P} + \mathsf{P}^2}}{\sigma^2} + \frac{\sqrt{\sigma^2\mathsf{P}_{\mathrm{FB}} + \mathsf{P}_{\mathrm{FB}}^2}}{\sigma_{\mathrm{FB}}^2} && (122)\\
&= \frac{\left(\sqrt{\mathsf{P} + \sigma^2} + \sqrt{\mathsf{P}}\right)^2}{2\sigma^2} + \frac{\left(\sqrt{\mathsf{P}_{\mathrm{FB}} + \sigma_{\mathrm{FB}}^2} + \sqrt{\mathsf{P}_{\mathrm{FB}}}\right)^2}{2\sigma_{\mathrm{FB}}^2} && (123)
\end{aligned}$$

with equality if $\mathsf{P}^{(0)} = \mathsf{P}^{(1)} = \mathsf{P}$ and $\mathsf{P}_{\mathrm{FB}}^{(0)} = \mathsf{P}_{\mathrm{FB}}^{(1)} = \mathsf{P}_{\mathrm{FB}}$. Here the inequality (120) follows from Jensen's inequality.

We conclude from (123) that the suboptimal choice (110) demonstrates that the error exponent is upper-bounded by

$$\frac{\left(\sqrt{\mathsf{P}+\sigma^2}+\sqrt{\mathsf{P}}\right)^2}{\sigma^2}+\frac{\left(\sqrt{\mathsf{P}_{\mathrm{FB}}+\sigma_{\mathrm{FB}}^2}+\sqrt{\mathsf{P}_{\mathrm{FB}}}\right)^2}{\sigma_{\mathrm{FB}}^2}. \tag{124}$$

Note that while the choice (110) is in general suboptimal, optimizing over $a_1$, $a_2$, $b_1$, and $b_2$ cannot yield any bound lower than half the bound of (124), because (101) and (102) imply that

$$a_1, a_2, b_1, b_2 \geq 1.$$

# 7 Achievability under Expected Power Constraints

We begin with the description of a "building block" comprising a transmission scheme upon which we shall build when coding under an expected power constraint on the feedback link.

## 7.1 A building block

**The Scheme:** The building block transmits a single bit, which we denote by $H$. Its parameters are the blocklength $n$, the transmitted average power on the forward link $\mathsf{P} > 0$, the noise variances $\sigma^2, \sigma_{\mathrm{FB}}^2 > 0$ on the forward and backward links, and the average transmitted power on the feedback link $\Delta > 0$. Note that $\Delta$ will not influence the error exponent of the building block. Consequently, when we later use the building block we shall typically choose $\Delta$ very small to save power. An additional parameter is some arbitrary (small) $0 < \delta < 1$.

The key elements of the building block are transmission, ACK/NACK, and retransmission. The transmission is based on antipodal signaling to send the bit over $n-1$ channel uses:

$$X_k = \begin{cases} +\sqrt{\mathsf{P}} & \text{if } H = 0, \\ -\sqrt{\mathsf{P}} & \text{if } H = 1, \end{cases} \quad k = 1, \ldots, n-1. \tag{125}$$

The first $n-2$ symbols sent by the receiver are zero

$$U_k = 0, \quad k = 1, \ldots, n-2. \tag{126}$$

Based on the received symbols $Y_1, \ldots, Y_{n-1}$, the receiver computes

$$S = \sum_{k=1}^{n-1} Y_k. \tag{127}$$

If

$$\left| \frac{1}{n-1} S \right| \le (1-\delta)\sqrt{\mathsf{P}},$$

then the decoder declares that a NACK event has occurred. Otherwise, it declares that an ACK event has occurred. It then uses $U_{n-1}$ to tell the transmitter which of these two events occurred:

$$U_{n-1} = \begin{cases} 0 & \text{if ACK} \\ \sqrt{\Delta/\mathsf{P}_0(\text{NACK})} & \text{if NACK.} \end{cases} \tag{128}$$

Notice that a NACK is a rare event of exponentially small probability:

$$\begin{aligned} \mathsf{P}_0(\text{NACK}) &= \mathsf{P}_0\left( \frac{S}{n-1} < (1-\delta)\sqrt{\mathsf{P}} \right) - \mathsf{P}_0\left( \frac{S}{n-1} < -(1-\delta)\sqrt{\mathsf{P}} \right) \\ &= \mathcal{Q}\left( \frac{\delta\sqrt{\mathsf{P}}}{\sigma/\sqrt{n-1}} \right) - \mathcal{Q}\left( \frac{(2-2\delta)\sqrt{\mathsf{P}}}{\sigma/\sqrt{n-1}} \right) \\ &\approx e^{-(n-1)\frac{\delta^2 \mathsf{P}}{2\sigma^2}}, \end{aligned} \tag{129}$$

and, by symmetry,

$$\mathsf{P}_1(\text{NACK}) = \mathsf{P}_0(\text{NACK}). \tag{130}$$

Based on $Z_{n-1}$ the transmitter guesses whether a NACK or ACK occurred. It does so by comparing $Z_{n-1}$ to some threshold $\Upsilon$. If $Z_{n-1}$ exceeds $\Upsilon$, then the transmitter guesses that a NACK was sent. We refer to this event as a "NACK". Otherwise, if $Z_{n-1}$ is smaller than $\Upsilon$, then the transmitter guesses that an ACK was sent and we refer to this as an "ACK" event. The threshold $\Upsilon$ is chosen so that both $\mathsf{P}_0(\text{"NACK"}|\text{ACK})$ and $\mathsf{P}_0(\text{"ACK"}|\text{NACK})$ decay in $n$ to zero faster than any exponential, i.e.,

$$-\lim_{n\to\infty} \frac{1}{n} \ln \mathsf{P}_0(\text{"NACK"}|\text{ACK}) = +\infty, \tag{131a}$$

$$-\lim_{n\to\infty} \frac{1}{n} \ln \mathsf{P}_0(\text{"ACK"}|\text{NACK}) = +\infty. \tag{131b}$$

In view of (128) and (129), this can be accomplished, for example, by choosing $\Upsilon = n$.

The time-$n$ transmitted symbol $X_n$ is now determined as follows. If "ACK", then the transmitter sends the zero symbol; otherwise it re-transmits $H$ using antipodal signaling with very large power. Thus,

$$X_n = \begin{cases} 0 & \text{if "ACK"}, \\ +\sqrt{\mathsf{P}/\mathsf{P}_0(\text{"NACK"})} & \text{if "NACK" and } H = 0, \\ -\sqrt{\mathsf{P}/\mathsf{P}_0(\text{"NACK"})} & \text{if "NACK" and } H = 1. \end{cases} \tag{132}$$

Note that "NACK" has an exponentially small probability. (This follows because NACK is a rare event (129) and because of (131).) Consequently, in (132) the symbol $X_n$ that is sent if "NACK" occurs have a huge magnitude so

$$-\lim_{n\to\infty} \frac{1}{n} \ln \mathsf{P}_0\big(Y_n < 0 | \text{"NACK"}\big) = +\infty, \tag{133a}$$

$$-\lim_{n\to\infty} \frac{1}{n} \ln \mathsf{P}_1\big(Y_n > 0 | \text{"NACK"}\big) = +\infty. \tag{133b}$$

We next consider how the receiver forms its guess $\hat{H}$ for $H$. If ACK, then the receiver guesses based on the sign of $S$. If NACK, then it guesses based on the sign of $Y_n$. Thus,

$$\hat{H} = \begin{cases} 0 & \text{if } \big(\text{ACK and } S > 0\big) \text{ or } \big(\text{NACK and } Y_n > 0\big), \\ 1 & \text{if } \big(\text{ACK and } S < 0\big) \text{ or } \big(\text{NACK and } Y_n < 0\big). \end{cases} \tag{134}$$

**Analysis:** We next analyze the probability of error. Since the scheme is completely symmetric, it suffices to compute the conditional probability of error conditional on $H = 0$, namely, $\mathsf{P}_0(\text{error})$. We express this probability as

$$\begin{aligned} \mathsf{P}_0(\text{error}) = {} & \overbrace{\mathsf{P}_0(\text{NACK})}^{\leq 1} \overbrace{\mathsf{P}_0(\text{"NACK"}|\text{NACK})}^{\leq 1} \overbrace{\mathsf{P}_0(\text{error}|\text{NACK, "NACK"})}^{\text{exceedingly small by (133a)}} \\ & + \overbrace{\mathsf{P}_0(\text{NACK})}^{\leq 1} \overbrace{\mathsf{P}_0(\text{"ACK"}|\text{NACK})}^{\text{exceedingly small by (131b)}} \overbrace{\mathsf{P}_0(\text{error}|\text{NACK, "ACK"})}^{\leq 1} \\ & + \overbrace{\mathsf{P}_0(\text{ACK})}^{\leq 1} \overbrace{\mathsf{P}_0(\text{"NACK"}|\text{ACK})}^{\text{exceedingly small by (131a)}} \overbrace{\mathsf{P}_0(\text{error}|\text{ACK, "NACK"})}^{\leq 1} \\ & + \overbrace{\mathsf{P}_0(\text{ACK})}^{\leq 1} \overbrace{\mathsf{P}_0(\text{"ACK"}|\text{ACK})}^{\leq 1} \mathsf{P}_0(\text{error}|\text{ACK, "ACK"}) \end{aligned}$$

$$\approx \mathsf{P}_0(\text{error}|\text{ACK}, \text{"ACK"})$$
$$= \mathsf{P}_0(\text{error}|\text{ACK}). \tag{135}$$

We next study $\mathsf{P}_0(\text{error}|\text{ACK})$:

$$\mathsf{P}_0(\text{error}|\text{ACK}) = \mathsf{P}_0\Big(S < 0 \Big| (n-1)^{-1}|S| > (1-\delta)\sqrt{\mathsf{P}}\Big) \tag{136}$$
$$= \frac{\mathsf{P}_0\Big((n-1)^{-1}S < -(1-\delta)\sqrt{\mathsf{P}}\Big)}{\mathsf{P}_0\Big((n-1)^{-1}|S| > (1-\delta)\sqrt{\mathsf{P}}\Big)} \tag{137}$$
$$= \frac{\mathcal{Q}\left(\frac{(2-\delta)\sqrt{\mathsf{P}}}{\sigma/\sqrt{n-1}}\right)}{\mathsf{P}_0\Big((n-1)^{-1}|S| > (1-\delta)\sqrt{\mathsf{P}}\Big)} \tag{138}$$
$$\approx \exp\left(-n\frac{(2-\delta)^2\mathsf{P}}{2\sigma^2}\right). \tag{139}$$

Since $\delta > 0$ can be chosen as small as we want, we conclude from (135) and (139) that the building block we described can achieve any error exponent smaller than

$$\frac{2\mathsf{P}}{\sigma^2}. \tag{140}$$

In fact, if we allow for $\delta$ to tends to zero very slowly with $n$, we can even achieve this exponent irrespective of how small $\Delta > 0$ is.

## 7.2 The Proposed Scheme and its Performance

**The scheme:** The proposed scheme has three phases: a transmission phase, an echo phase, and a re-transmission phase. To simplify the exposition we shall assume that the blocklength $n$ is odd and define

$$\nu = \frac{1}{2}(n-1). \tag{141}$$

(To code for an even blocklength $n$ we can use the proposed scheme for $n-1$ and then append a zero symbol to the transmission.) Fix some $0 < \Delta < \min\{\mathsf{P}, \mathsf{P}_{\text{FB}}\}$ (later to be chosen arbitrarily small).

In the *transmission phase* we use the building block of Section 7.1 to transmit $H$ in $\nu$ channel uses using power $2\mathsf{P} - \Delta$ on the forward link and power $\Delta$ on the feedback link. We denote the receiver's guess of $H$ after this phase by $H'$. By (140) and (141) we have

$$\mathsf{P}_0\bigl(H' \neq H\bigr) = \mathsf{P}_1\bigl(H' \neq H\bigr) \approx \exp\Bigl(-n\frac{2\mathsf{P}-\Delta}{\sigma^2}\Bigr). \tag{142}$$

In the *echo phase* we use the building block of Section 7.1 to send $H'$ *from the receiver to the transmitter*. We thus think of the original feedback channel from the receiver to the transmitter as the forward channel in the building block, and we think of the original forward channel from the transmitter to the receiver as the feedback channel in the building block. The receiver uses power $2\mathsf{P}_{\mathrm{FB}} - \Delta$ and the transmitter uses power $\Delta$. We denote the transmitter's guess for $H'$ at the end of this phase by $H''$. Substituting in (140) $2\mathsf{P}_{\mathrm{FB}} - \Delta$ for $\mathsf{P}$ and $\sigma^2_{\mathrm{FB}}$ for $\sigma^2$ we obtain that

$$\mathsf{P}_0\bigl(H'' \neq H'\bigr) = \mathsf{P}_1\bigl(H'' \neq H'\bigr) \approx \exp\Bigl(-n\frac{2\mathsf{P}_{\mathrm{FB}}-\Delta}{\sigma^2_{\mathrm{FB}}}\Bigr). \tag{143}$$

Note that by (142) and (143)

$$\begin{aligned}\mathsf{P}_0\bigl(H''=0, H'=1\bigr) &= \mathsf{P}_1\bigl(H''=1, H'=0\bigr)\\ &\approx \exp\biggl(-n\Bigl(\frac{2\mathsf{P}-\Delta}{\sigma^2} + \frac{2\mathsf{P}_{\mathrm{FB}}-\Delta}{\sigma^2_{\mathrm{FB}}}\Bigr)\biggr).\end{aligned} \tag{144}$$

In the final re-transmission phase which comprises of one channel use, the transmitter compares $H$ and $H''$. If they are equal, it sends the zero symbol. Otherwise it re-transmits $H$ using antipodal signaling with amplitude $\sqrt{\mathsf{P}/\mathsf{P}_0(H'' \neq H)}$:

$$X_n = \begin{cases} 0 & \text{if } H'' = H,\\ +\sqrt{\mathsf{P}/\mathsf{P}_0(H'' \neq H)} & \text{if } (H, H'') = (0,1),\\ -\sqrt{\mathsf{P}/\mathsf{P}_0(H'' \neq H)} & \text{if } (H, H'') = (1,0).\end{cases} \tag{145}$$

At the end of the re-transmission phase, the receiver makes its final guess $\hat{H}$ of $H$. It does so by comparing $Y_n$ to a threshold $\Upsilon$:

$$\hat{H} = \begin{cases} 0 & \text{if } Y_n > \Upsilon,\\ H' & \text{if } |Y_n| \leq \Upsilon,\\ 1 & \text{if } Y_n < -\Upsilon.\end{cases} \tag{146}$$

Notice that by (144) the amplitude $\sqrt{\mathsf{P}/\mathsf{P}_0(H'' \neq H)}$ is exponentially large, so there is no difficulty finding a threshold $\Upsilon$ (e.g., $\Upsilon = n$) such that

$$-\lim_{n\to\infty} \frac{1}{n} \ln \Pr\Big[|Y_n| > \Upsilon \,\Big|\, X_n = 0\Big] = +\infty, \tag{147a}$$

$$-\lim_{n\to\infty} \frac{1}{n} \ln \Pr\Big[Y_n \le \Upsilon \,\Big|\, X_n = \sqrt{\mathsf{P}/\mathsf{P}_0(H'' \neq H)}\Big] = +\infty, \tag{147b}$$

$$-\lim_{n\to\infty} \frac{1}{n} \ln \Pr\Big[Y_n \ge -\Upsilon \,\Big|\, X_n = -\sqrt{\mathsf{P}/\mathsf{P}_0(H'' \neq H)}\Big] = +\infty. \tag{147c}$$

**Analysis:** We shall analyze the probability of error conditional on $H = 0$. It is identical to the one given $H = 1$.

$$\begin{aligned}
\mathsf{P}_0(\text{error}) &= \overbrace{\mathsf{P}_0\big(\text{error} \mid H'' = 0,\ H' = 0\big)}^{\text{negligible by (147a)}} \overbrace{\mathsf{P}_0\big(H'' = 0,\ H' = 0\big)}^{\le 1} \\
&\quad + \overbrace{\mathsf{P}_0\big(\text{error}|H'' = 0,\ H' = 1\big)}^{\le 1} \mathsf{P}_0\big(H'' = 0,\ H' = 1\big) \\
&\quad + \overbrace{\mathsf{P}_0\big(\text{error}|H'' = 1,\ H' = 0\big)}^{\text{negligible by (147b)}} \overbrace{\mathsf{P}_0\big(H'' = 1,\ H' = 0\big)}^{\le 1} \\
&\quad + \overbrace{\mathsf{P}_0\big(\text{error}|H'' = 1,\ H' = 1\big)}^{\text{negligible by (147b)}} \overbrace{\mathsf{P}_0\big(H'' = 1,\ H' = 1\big)}^{\le 1} \\
&\approx \mathsf{P}_0\big(H'' = 0,\ H' = 1\big) \\
&\approx \exp\left(-n\Big(\frac{2\mathsf{P} - \Delta}{\sigma^2} + \frac{2\mathsf{P}_{\text{FB}} - \Delta}{\sigma^2_{\text{FB}}}\Big)\right),
\end{aligned} \tag{148}$$

where the last approximation follows from (144). By symmetry, $\mathsf{P}_1(\text{error}) = \mathsf{P}_0(\text{error})$, so (148) implies that the error exponent

$$\frac{2\mathsf{P} - \Delta}{\sigma^2} + \frac{2\mathsf{P}_{\text{FB}} - \Delta}{\sigma^2_{\text{FB}}} \tag{149}$$

is achievable. Since, $\Delta$ can be chosen arbitrarily small, it follows that any exponent smaller than

$$\frac{2\mathsf{P}}{\sigma^2} + \frac{2\mathsf{P}_{\text{FB}}}{\sigma^2_{\text{FB}}} \tag{150}$$

is achievable. In fact, if we let $\Delta$ tend to zero very slowly with $n$, we can achieve the error exponent (150).

# 8 Summary and Possible Extensions

We presented results on the best achievable error exponents in transmitting a bit over the Gaussian channel with an active noisy Gaussian feedback link. (In the case of an almost sure block power constraint on the feedback link we also obtained an upper bound on the error exponent for arbitrary rates of communication; see (51).) We have shown that even if both the forward link and the feedback link are subjected to expected block power constraints, the best error exponents are finite. Roughly speaking—irrespective of the nature of the feedback power constraint—the best error exponent is roughly proportional to the larger of the signal-to-noise ratio on the forward link $\mathsf{P}/\sigma^2$ and the signal-to-noise ratio on the feedback channel $\mathsf{P}_{\mathrm{FB}}/\sigma^2_{\mathrm{FB}}$. In this very rough sense, active feedback is not much different from passive symbol-by-symbol feedback [2].

However, a more careful analysis based on our previous result [1, 2] shows that the best error exponent for two messages with passive (symbol-by-symbol) feedback is upper-bounded by

$$\frac{1}{2}\left(\frac{\mathsf{P}}{\sigma^2} + \frac{\mathsf{P}}{\mathsf{P}+\sigma^2}\cdot\frac{\mathsf{P}_{\mathrm{FB}}}{\sigma^2_{\mathrm{FB}}}\right), \tag{151}$$

which can be further upper-bounded by

$$\frac{1}{2}\left(\frac{\mathsf{P}}{\sigma^2} + \frac{\mathsf{P}_{\mathrm{FB}}}{\sigma^2_{\mathrm{FB}}}\right). \tag{152}$$

On the other hand, an achievable error exponent (150) for an active feedback with the same feedback signal-to-noise ratio is

$$2\left(\frac{\mathsf{P}}{\sigma^2} + \frac{\mathsf{P}_{\mathrm{FB}}}{\sigma^2_{\mathrm{FB}}}\right).$$

Hence, the freedom to code over the feedback link can at least quadruple the error exponent of binary communication. It would be interesting to see how much active feedback gains over passive feedback for a positive rate $R$.

While our focus has been on Gaussian channels with Gaussian feedback channels, some of our techniques are more general. For example, consider a setting where the forward and feedback channels are binary symmetric channels (BSCs) with crossover probabilities $\epsilon, \epsilon_{\mathrm{FB}} \leq 1/2$. In this case we can apply the technique of Section 4 (with the sets $\mathcal{A}$ and $\mathcal{B}$ being $\{0,1\}^n$) to obtain that the reliability function with noisy active feedback cannot exceed

$$\ln \frac{1-\boldsymbol{\epsilon}_{\mathrm{FB}}}{\boldsymbol{\epsilon}_{\mathrm{FB}}} + E_{\mathrm{NoFB}}(R;\boldsymbol{\epsilon}), \tag{153}$$

where $E_{\mathrm{NoFB}}(R;\boldsymbol{\epsilon})$ is the reliability function of the BSC of crossover probability $\boldsymbol{\epsilon}$.

In some cases, particularly when the feedback channel is very noisy, this bound can be tighter than the trivial bound that bounds the reliability function by that with perfect feedback and bounds the latter by the best two-codeword error exponent

$$\frac{1}{2} \ln \frac{1}{4\boldsymbol{\epsilon}(1-\boldsymbol{\epsilon})}. \tag{154}$$

(The fact that feedback does not improve the best two-codeword error exponent on a discrete memoryless channel appears in the Ph.D. thesis of Berlekamp [10] who attributes this result to Gallager and Shannon.)

The bound in (153) complements the recent work of Burnashev and Yamamoto [11] on the reliability function of the binary symmetric channel with a passive binary symmetric feedback link. (Upper bounds on the latter reliability function can be derived using techniques similar to those we used in [2].)